\documentclass[apj,twocolumn]{aastex63}
\usepackage{xspace}

\usepackage[T1]{fontenc}
\usepackage{ae,aecompl}

\usepackage{graphicx}   
\usepackage{amsmath}    
\usepackage{amssymb}    
\usepackage{url}

\usepackage[normalem]{ulem}
\usepackage{color}
\usepackage{bm}
\usepackage{booktabs} 
\usepackage{comment}

\def\mlt/{\ifmmode \alpha_{\text{MLT}}\else $\alpha_{\text{MLT}}$\fi}

\newcommand{\ctwelvecthirteen}{$^{12}$C/$^{13}$C\xspace}
\newcommand{\hethree}{$^3$He\xspace}
\newcommand{\msun}{$\mathrm{M_{\odot}}$}
\begin{document}

\title{Is Thermohaline Mixing the Full Story? Evidence for Separate Mixing Events near the Red Giant Branch Bump}
\shorttitle{Is It Thermohaline?}

\author[0000-0002-4818-7885]{Jamie Tayar}
\altaffiliation{These authors contributed equally to this work}
\affiliation{Department of Astronomy, University of Florida, Bryant Space Science Center, Stadium Road, Gainesville, FL 32611, USA }
\affiliation{Institute for Astronomy, University of Hawai‘i at Mānoa, 2680 Woodlawn Drive, Honolulu, HI 96822, USA}
\affiliation{NASA Hubble Fellow}
\author[0000-0002-8717-127X]{Meridith Joyce}
\altaffiliation{These authors contributed equally to this work}
\affiliation{Space Telescope Science Institute, 3700 San Martin Dr, Baltimore, MD 21202, USA}
\affiliation{Kavli Institute for Theoretical Physics, University of Santa Barbara, California, 93106, USA}

\shortauthors{Tayar \& Joyce}

\correspondingauthor{Jamie Tayar}
\email{jtayar@ufl.edu}
\correspondingauthor{Meridith Joyce}
\email{mjoyce@stsci.edu}

\label{firstpage}

\date{Accepted XXX. Received YYY; in original form ZZZ}

\begin{abstract}
The abundances of mixing--sensitive elements including lithium, [C/N], and \ctwelvecthirteen\ are known to change near the red giant branch bump. The explanation most often offered for these alterations is double diffusive thermohaline mixing in the stellar interior. In this analysis, we investigate the ability of thermohaline mixing to explain the observed timing of these chemical depletion events.
Recent observational measurements of lithium and [C/N] show that the abundance of lithium decreases before the abundance of [C/N], whereas numerical simulations of the propagation of the thermohaline mixing region {computed with MESA} show that the synthetic abundances drop simultaneously. We therefore conclude that thermohaline mixing alone cannot explain the distinct events of lithium depletion and [C/N] depletion, as the simultaneity predicted by simulations is not consistent with the observation of separate drops. We thus invite more sophisticated theoretical explanations for the observed temporal separation of these chemical depletion episodes as well as more extensive observational explorations across a range of masses and metallicities.
\end{abstract}


\section{Introduction}
As low-mass stars evolve into red giants, the surface abundances of some elements change in ways that depend on the internal stellar structure. On the early giant branch, the surface convection zone is deepening, diluting the abundance of delicate elements like lithium and beryllium that only survived the main sequence in a narrow region near the surface of the star \citep{Sweigart1989, Charbonnel1994, BoothroydSackmann1999, Boesgaard2020}. As the surface convection zone deepens to its greatest extent in mass, it penetrates into regions that have previously undergone nuclear processing and cycles the processed material towards the surface during the first dredge up, leaving behind a chemical discontinuity (see, e.g., \citealt{iben1964, Lambert1981, Joyce2015} and references therein). Due to the nuclear fusion rates in the carbon, nitrogen, oxygen (CNO) cycle and the temperature dependence of this depth, this process will alter the [C/N] and \ctwelvecthirteen\ ratios in a mass--dependent way \citep{Iben1967, Gratton2000, MasseronGilmore2015, Martig2016, Ness2016}. In standard stellar models without additional mixing or physics, the abundances at the surface of the star do not change between this point and the end of the red giant branch \citep{Chaname2005, Palacios2006}.  

After the first dredge--up, the surface convection zone recedes in mass coordinate as the hydrogen burning shell advances. Eventually the hydrogen burning shell crosses the residual chemical discontinuity 
and encounters the mean molecular weight gradient left behind by the deepest extent of the surface convection zone \citep{kippy, FusiPecci1990, Charbonnel1994, Joyce2015}. At this point, the structure readjusts, the core contracts and the luminosity decreases. In observed stellar populations, this is seen as an over-density of stars and identified as the red giant branch bump (RGBB). In stellar evolution models, the exact location of the RGBB depends on choices about the envelope mixing and undershoot parameters, which will likewise affect the structure and temperature at the base of the convection zone \citep{Joyce2015, Khan2018}. 

Observational work has noted that several abundances, including lithium, [C/N], and \ctwelvecthirteen\,  change for stars near the red giant branch bump, particularly at low metallicity  \citep[see e.g.][]{Carbon1982, Pilachowski1986, Kraft1994,Gratton2000}. More recent work has made it clear that the amount of extra mixing in these stars is strongly metallicity dependent \citep{Shetrone2019} and also depends on stellar mass \citep{Magrini2021a}. 

Great theoretical efforts have been made to try to infer the physical cause of this extra mixing. Given that the extra mixing occurs in a region where the mean molecular weight gradient is inverted from \hethree\ burning \citep{Eggleton2006}, authors including \citet{CharbonnelZahn2007} pointed out that the region should be unstable to double diffusive instabilities, also known as thermohaline mixing. While there were challenges to this theory \citep{Denissenkov2010, Denissenkov2011, Traxler2011, Brown2013, SenguptaGaraud2018, Garaud2019} on the grounds that the amount of mixing needed not did not quantitatively match the expectations given the fluid parameters, most authors have worked under the assumption that the extra mixing of all elements on the upper giant branch are somehow related to thermohaline mixing \citep[e.g.][]{Kirby2016, Charbonnel2020, Magrini2021c}.

Previous theoretical explorations have indicated that the amount of extra mixing in this regime should depend on stellar mass and metallicity \citep[e.g.][]{CharbonnelLagarde2010,Lagarde2017} and that the relevant process should affect several mixing tracers, including \ctwelvecthirteen, [C/Fe], [N/Fe], and $^7$Li, at approximately the same time above the red giant branch bump \citep{Stancliffe2009, CharbonnelLagarde2010}. Predicting the exact amount of mixing or the resulting abundances is extremely challenging, however, as the process can be affected by rotation \citep{Lagarde2011}, magnetism \citep{CharbonnelZahn2007b}, dynamical shear \citep{CantielloLanger2010}, numerical model choices \citep{Lattanzio2015}, and other complications, as well as the detailed interactions, rather than just additions, of these processes \citep{Maeder2013, SenguptaGaraud2018}. Further, the estimated degree of impact of any one of these processes likewise depends on the stellar properties, as has been discussed extensively \citep[see e.g.][]{CantielloLanger2010, Lagarde2011}. However, by making reasonable physical assumptions, several authors have been able to construct predictions for the mixing that at least roughly match the observations \citep[e.g.][]{Lagarde2019, Henkel2017, Henkel2018, Magrini2021a}.
 
In the following manuscript, we present observational and theoretical evidence for distinct mixing causes for the depletion of lithium and [C/N], from which it follows that thermohaline mixing is not sufficient explanation for all of the extra mixing that occurs near the red giant branch bump. We first present simulations showing that the connection of the thermohaline mixing region with the surface convection zone corresponds to simultaneous depletion of both lithium and [C/N]. We follow with the presentation of observational data demonstrating that {the abundance decreases in Li and [C/N] near the red giant branch bump are temporally distinct, with}
the depletion of lithium occurring earlier in a star's evolution than the depletion of [C/N] by more than 0.1 dex. {We thus demonstrate that a theoretical framework ascribing these events to thermohaline mixing alone is inconsistent with the observations.}


\section{Theoretical Exploration and Simulations}
Using the Modules for Experiments in Stellar Astrophysics (MESA; stable release version 15140) stellar evolution program, we evolve high structural and temporal resolution simulations of the interior mixing profiles of model stars in the approximate mass ($0.8-1.2 M_{\odot}$) and metallicity regime ($-2 \le$ [Fe/H] $\le 0$ dex, or $0.0007 \le Z \le$ 0.02) of the observed sample \citep{paxton2011, paxton2013, paxton2015, Paxton2018, paxton2019}. We use the solar mixture of \citet{GrevesseSauval1998} and OPAL opacities, metallicities corresponding to solar ($Z=0.02$) and [Fe/H]$=-1.2$ ($Z=0.0009$), the mixing length theory (MLT) implementation of \citet{CoxGiuli1968} 
with $\alpha_{\text{MLT}}=1.6$, the Eddington T-$\tau$ grey model atmosphere for surface boundary conditions, the \verb|pp_extras.net| nuclear reaction network, and the thermohaline mixing prescription of \citet{Kippenhahn1980} based on the analysis of \citet{Ulrich1972}
with a thermohaline efficiency coefficient $\alpha_{\text{th}}=2$ \citep{ Eddington1916, BoehmVitense1958, Cox1980, IglesiasRogers1996}.The choice to set $\alpha_{\text{th}}$ near unity was motivated by recent 2D and 3D hydrodynamic simulations showing that much lower values (e.g. $\alpha_{\text{th}} = 1$ vs $\alpha_{\text{th}}= 667$) are appropriate for stellar conditions as compared to salt fingers \citep{Denissenkov2010, Traxler2011, Brown2013}. 
A value of $\alpha_{\text{th}}=2$ specifically was chosen for consistency with Section 4.2 of MESA Instrument Paper II, in which MESA's thermohaline capabilities are described in more detail, and the associated \verb|test_suite| case \citep{paxton2013}. Using models described in \citet{Fraser2022}, we find that variations of $\alpha_{\text{th}}$ between 0.1 and 700 have no effect on our conclusions.
\setbox17\hbox{\verb|test_suite|}%
The Ledoux criterion for convective stability is used throughout \citep{Ledoux1951,Anders2022}. We improve MESA's default structural and temporal resolutions (\verb|mesh_delta_coeff| and \verb|time_delta_coeff|) by a factor of 10 to 100 in our calculations\footnote{Full details of our configuration will be made publicly available on Zenodo upon publication. A basic template for simulating thermohaline mixing is freely available in the MESA \copy17.}. In physical units, our time steps in this regime correspond to 21,000 years---reduced to as low as 3000 years in convergence investigations---and our spatial resolution corresponds to roughly 5500 zones. These are above the thresholds necessary to resolve the relevant changes due to thermohaline mixing according to, for example, \citet{Lattanzio2015}.
We note, however, that while even higher resolutions might be necessary to argue for particular values of calibrated parameters (for example, the mixing efficiency); this is not necessary in our case, and we propose no such values. It is sufficient for our argument to demonstrate that neither time step nor structural resolution affect the direction of propagation of the thermohaline front or the simultaneity of dilution events. This is indeed the case (see, e.g. Fraser et al. 2022, which includes extensive resolution tests using an identical modeling configuration).

\begin{figure}
    \centering
    \includegraphics[width=\columnwidth]{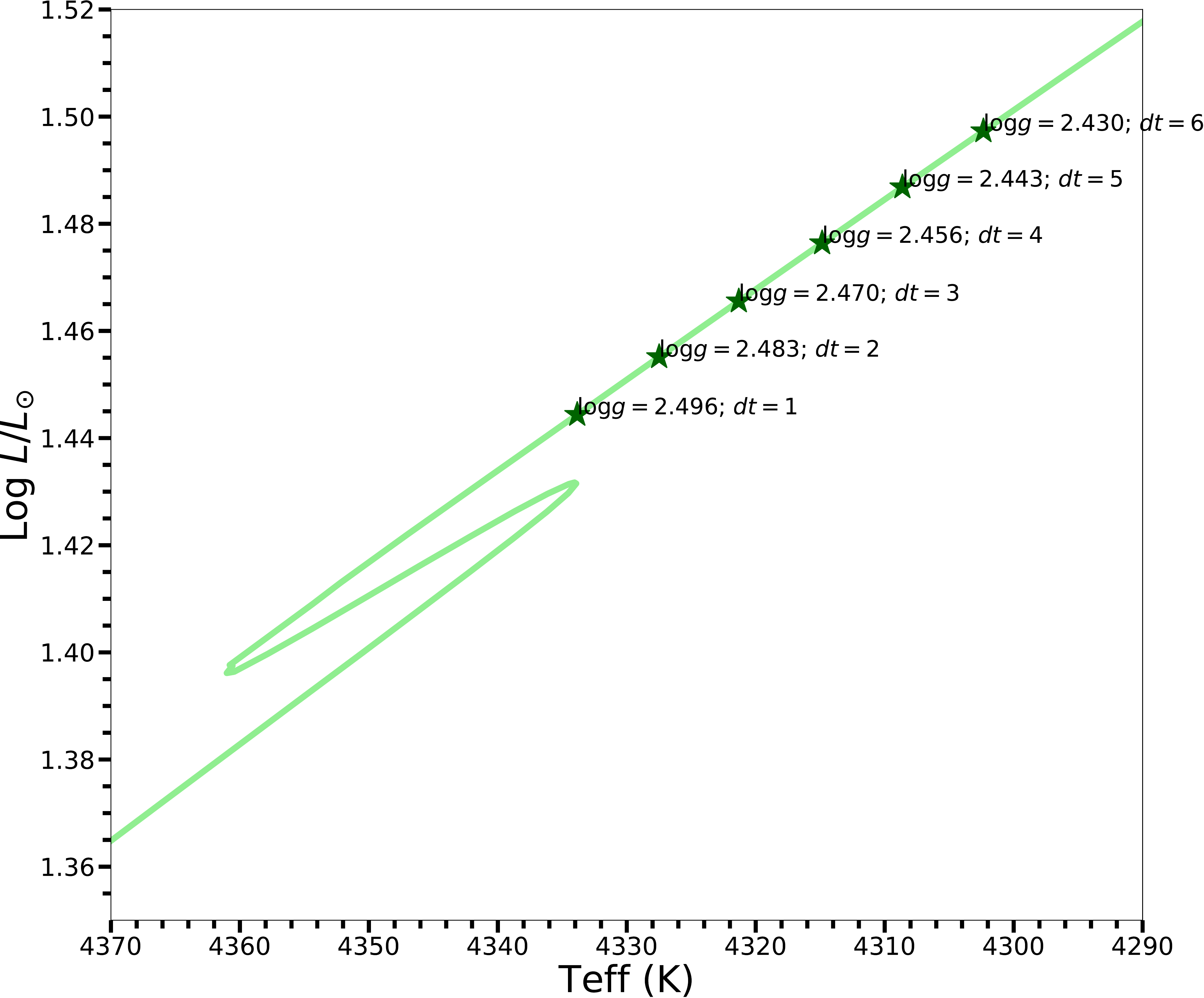}
    \caption{The evolutionary track for our canonical thermohaline propagation simulation { ($1 M_{\odot}$, solar metallicity $Z=0.020$)} is shown, zoomed in on the RGBB. Surface gravity is indicated at six time steps, denoted $dt$. These index the evolutionary times at which the subsequent stellar interior snapshots are taken. The time step indices are revised to begin at one for the purpose of demonstration only.}
    \label{fig:HR}
\end{figure}

The models are terminated at roughly $\log g \le 1$ on the upper red giant branch, after the RGBB. Figure \ref{fig:HR} shows an evolutionary track from one such simulation, centered on the evolutionary regime where thermohaline mixing is hypothesized to be important. Locations in the HR diagram and values of $\log g$ are indicated on the track for six time steps $dt$ corresponding to the structural snapshots presented subsequently. These are indexed beginning with one for convenience; in reality, there are thousands.

\begin{figure*}
\begin{center}
\includegraphics[width=\columnwidth]{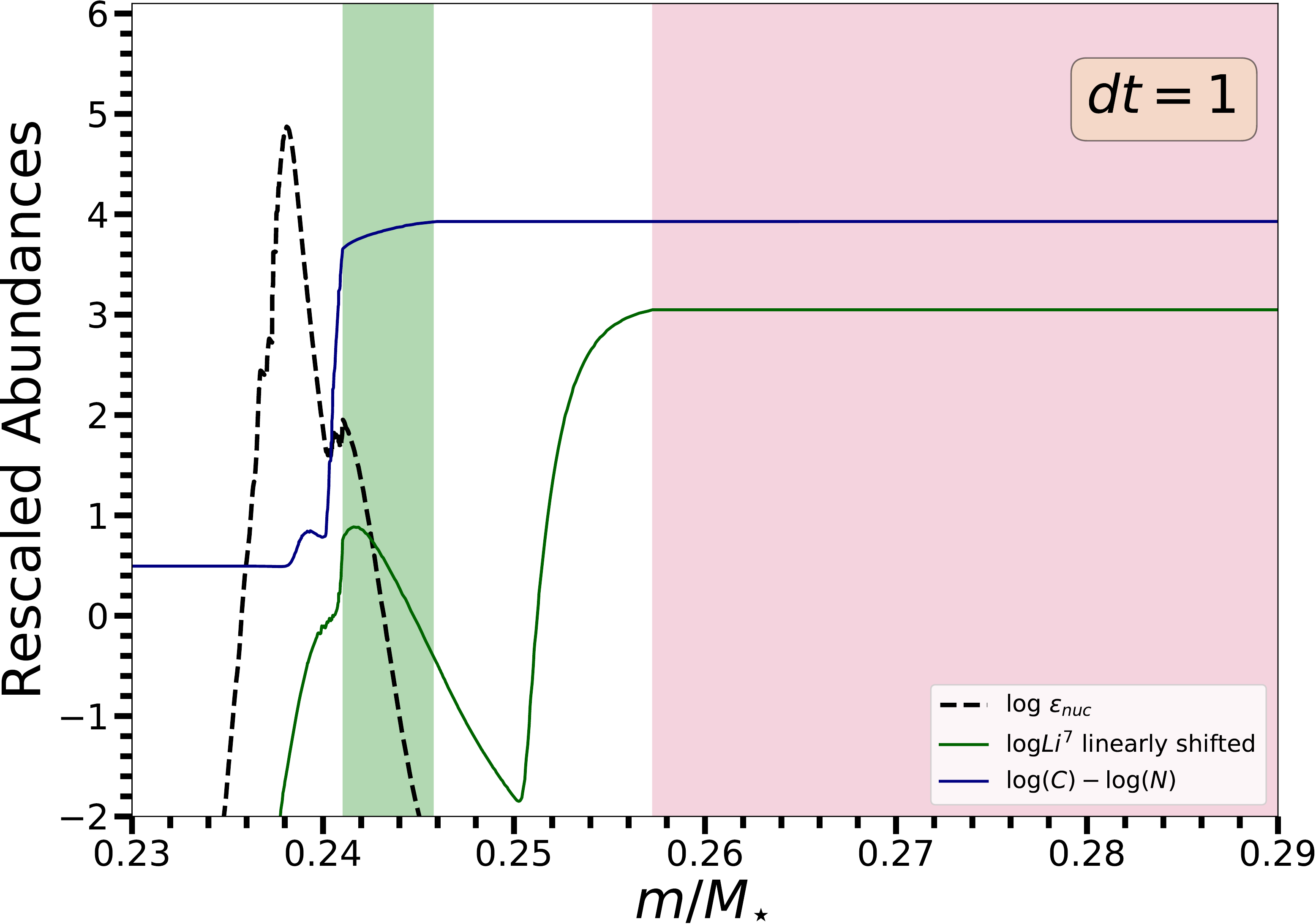}
\includegraphics[width=\columnwidth]{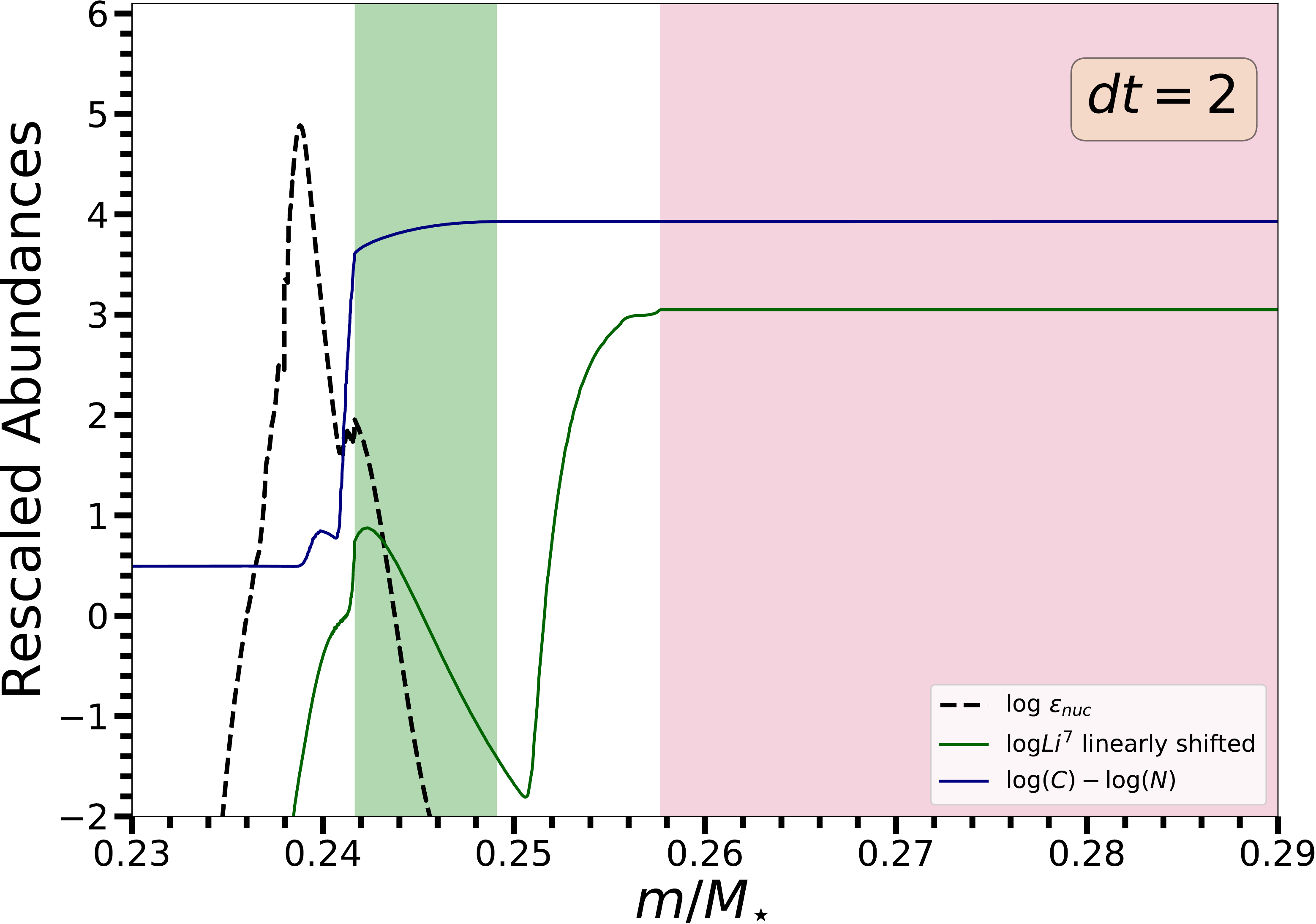}
\includegraphics[width=\columnwidth]{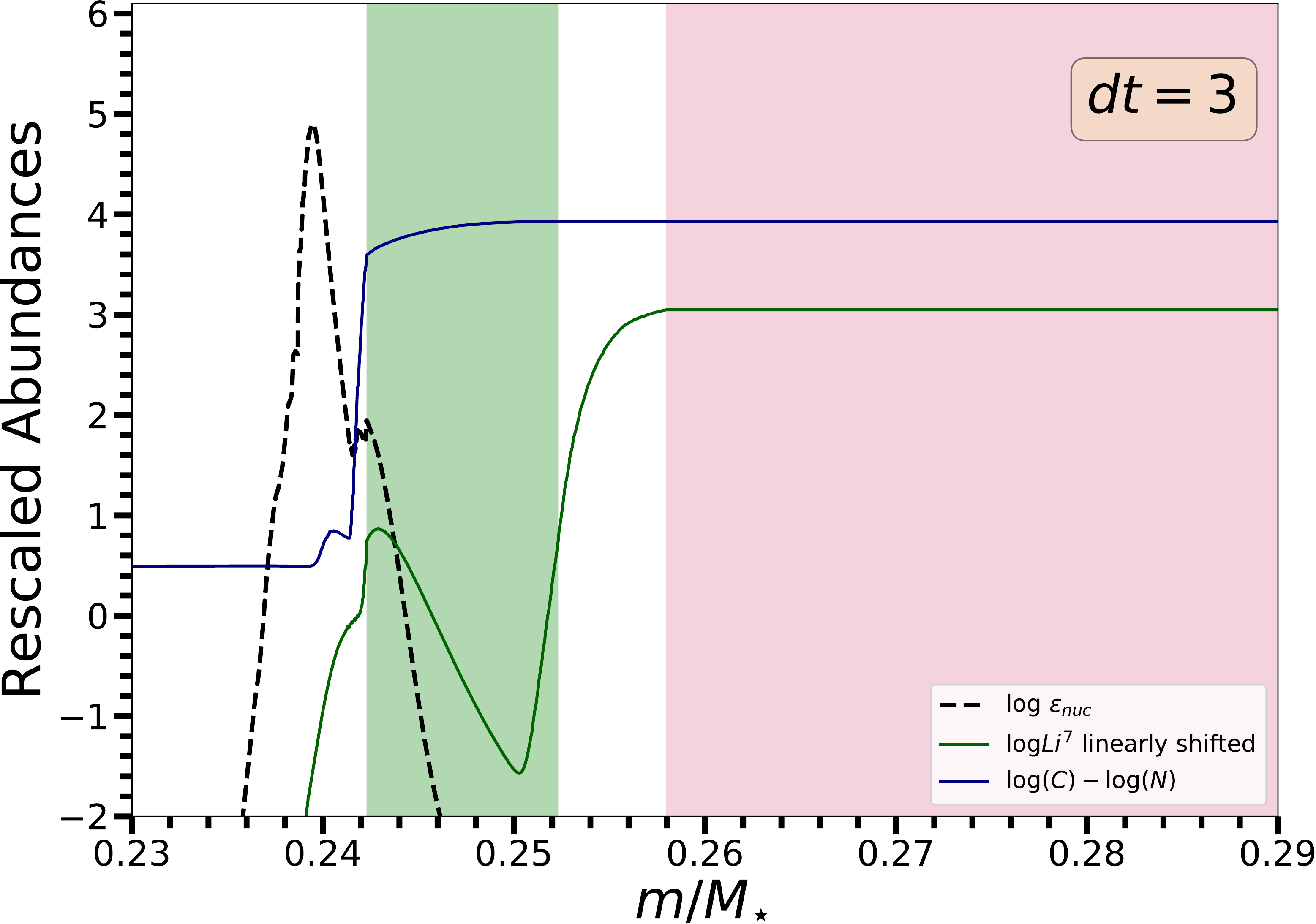}
\includegraphics[width=\columnwidth]{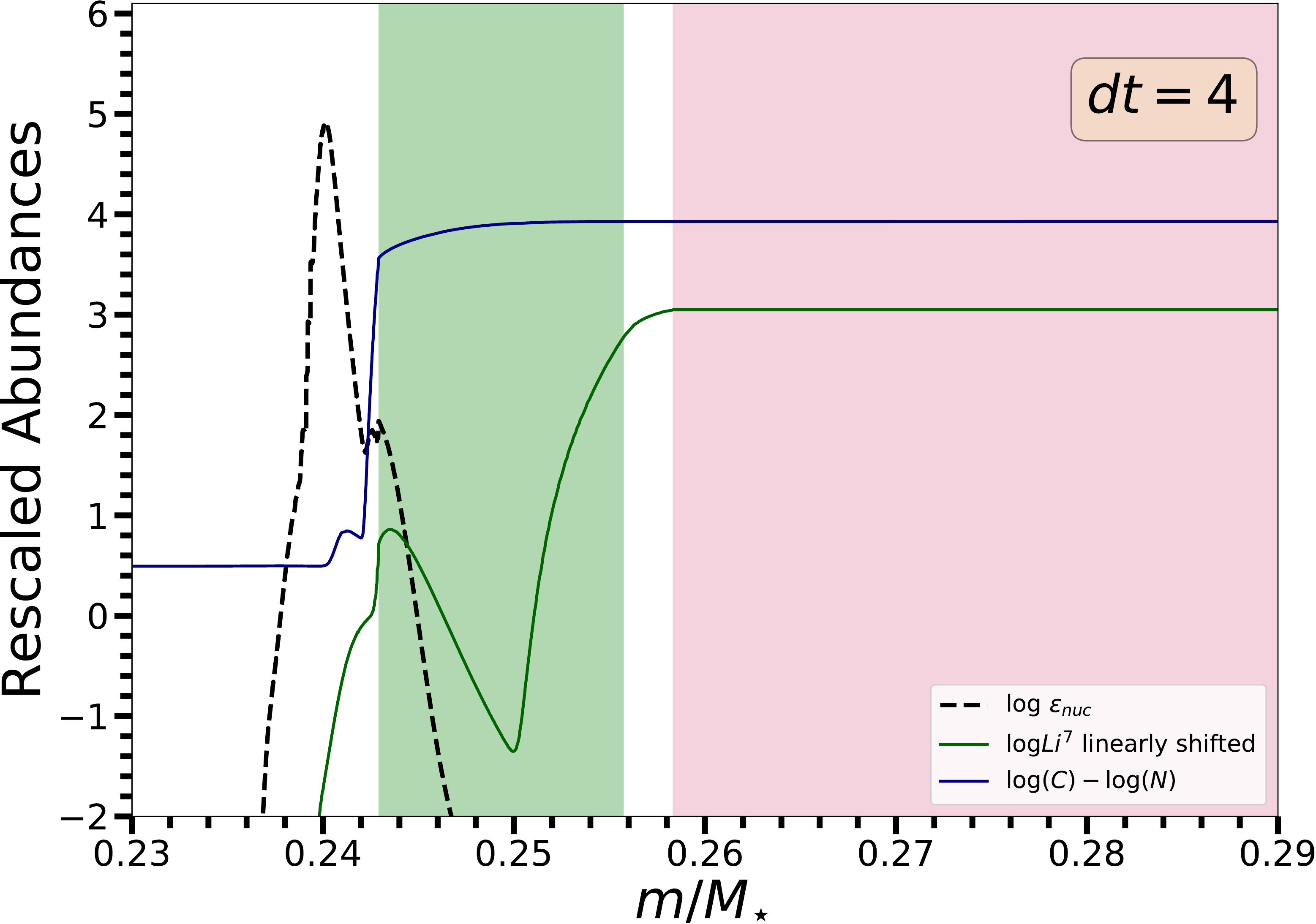}
\includegraphics[width=\columnwidth]{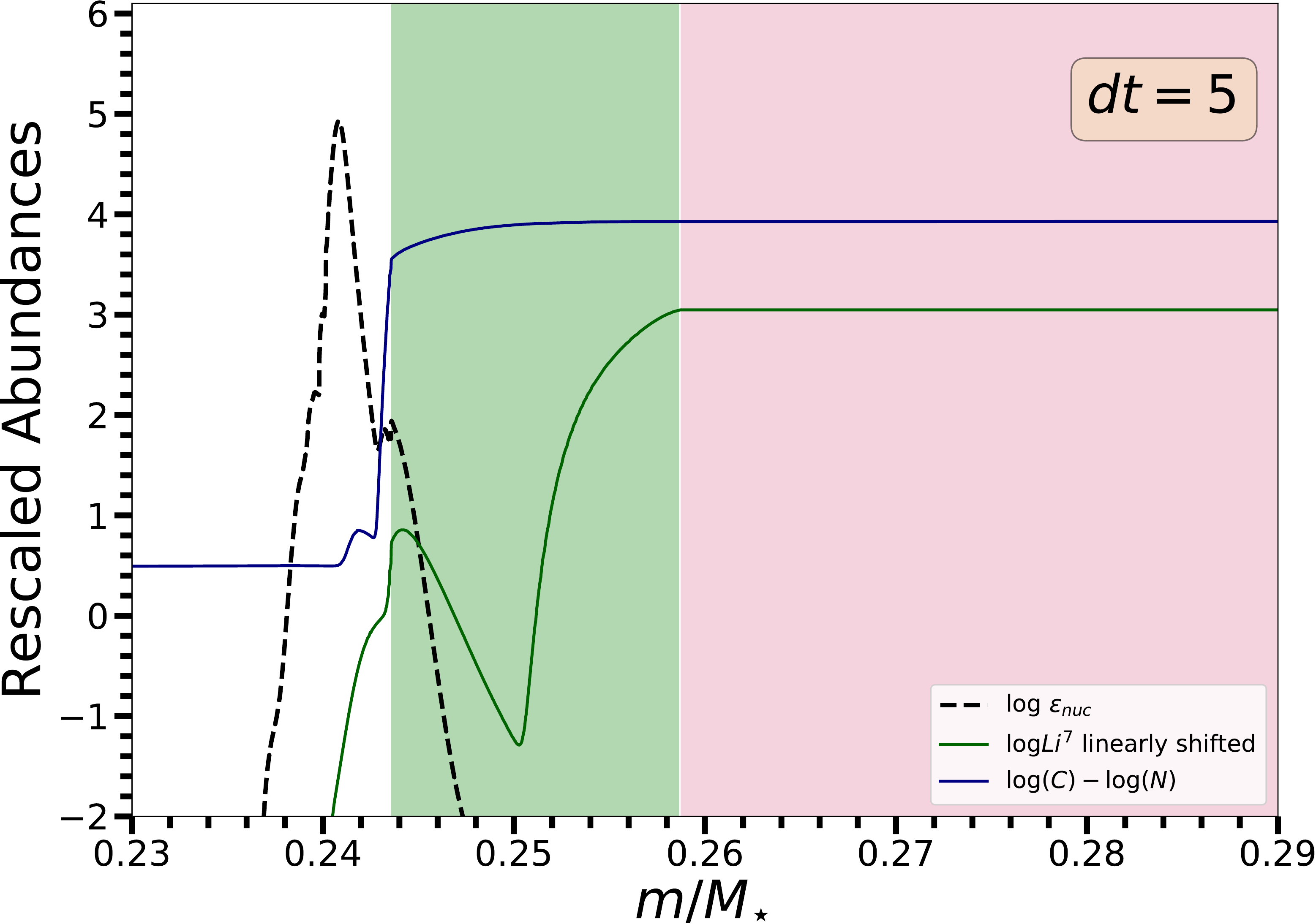}
\includegraphics[width=\columnwidth]{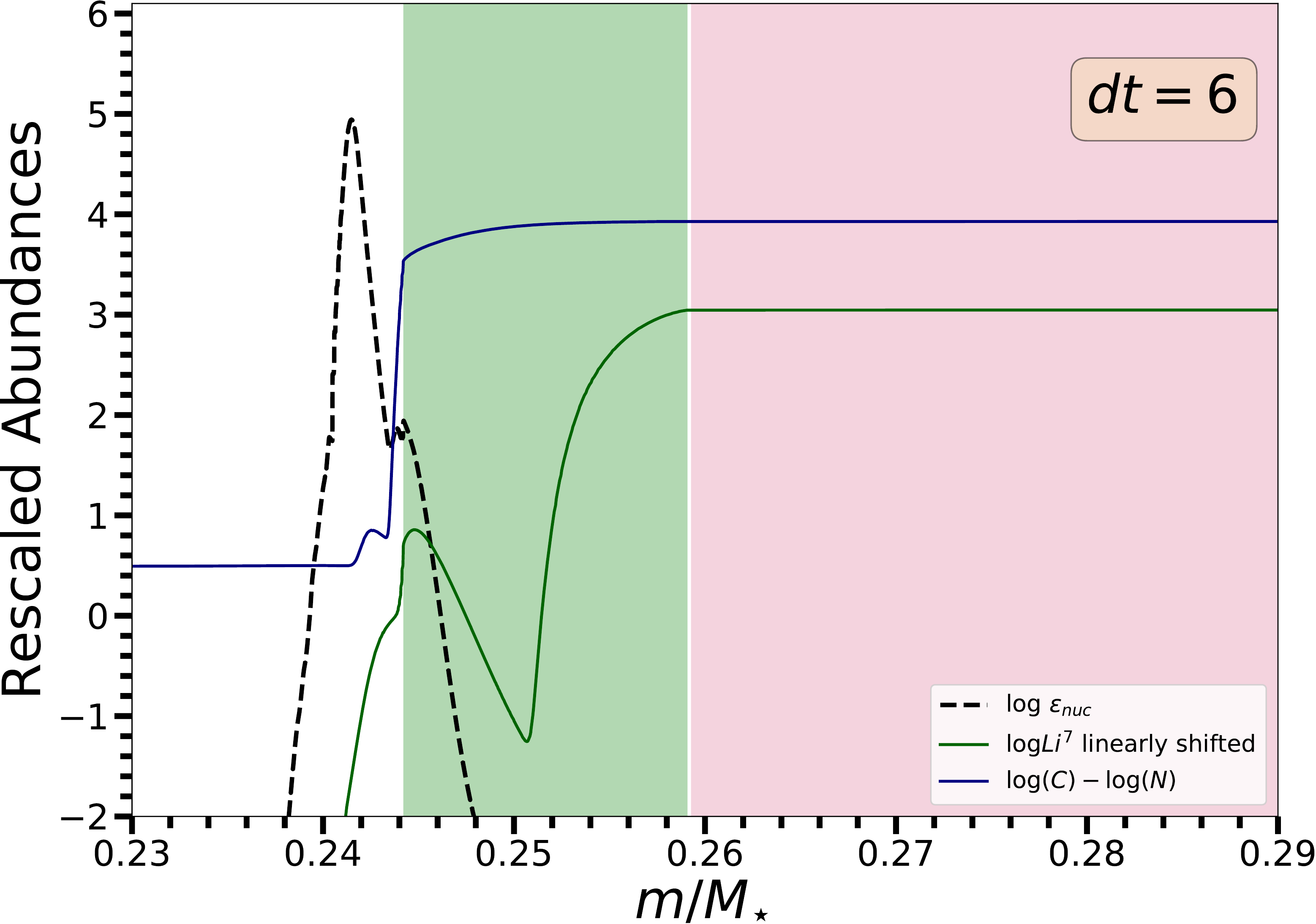}
\caption{Structural snapshots at evolutionary time steps for the $1 M_{\odot}$, solar composition model, as indicated in Figure \ref{fig:HR}, are shown. As the model steps through time (marked in the upper right corner) the nuclear burning shell (black dashed line) moves outwards, as the surface convection zone's (pink shaded region) size is reduced. The thermohaline region (green shaded region) grows from the hydrogen burning shell until it meets the base of the convective region. The abundance of lithium (green line) and the ratio of carbon to nitrogen (blue line) are also shown, with units rescaled to fit within the plot. In this simulation, the lithium depletion region and the {(C/N)} depleted region are connected to the surface convection zone by the thermohaline front at the same time. }
\label{fig:theory}
\end{center}
\end{figure*}

\begin{figure*}
\begin{center}
\includegraphics[width=\columnwidth]{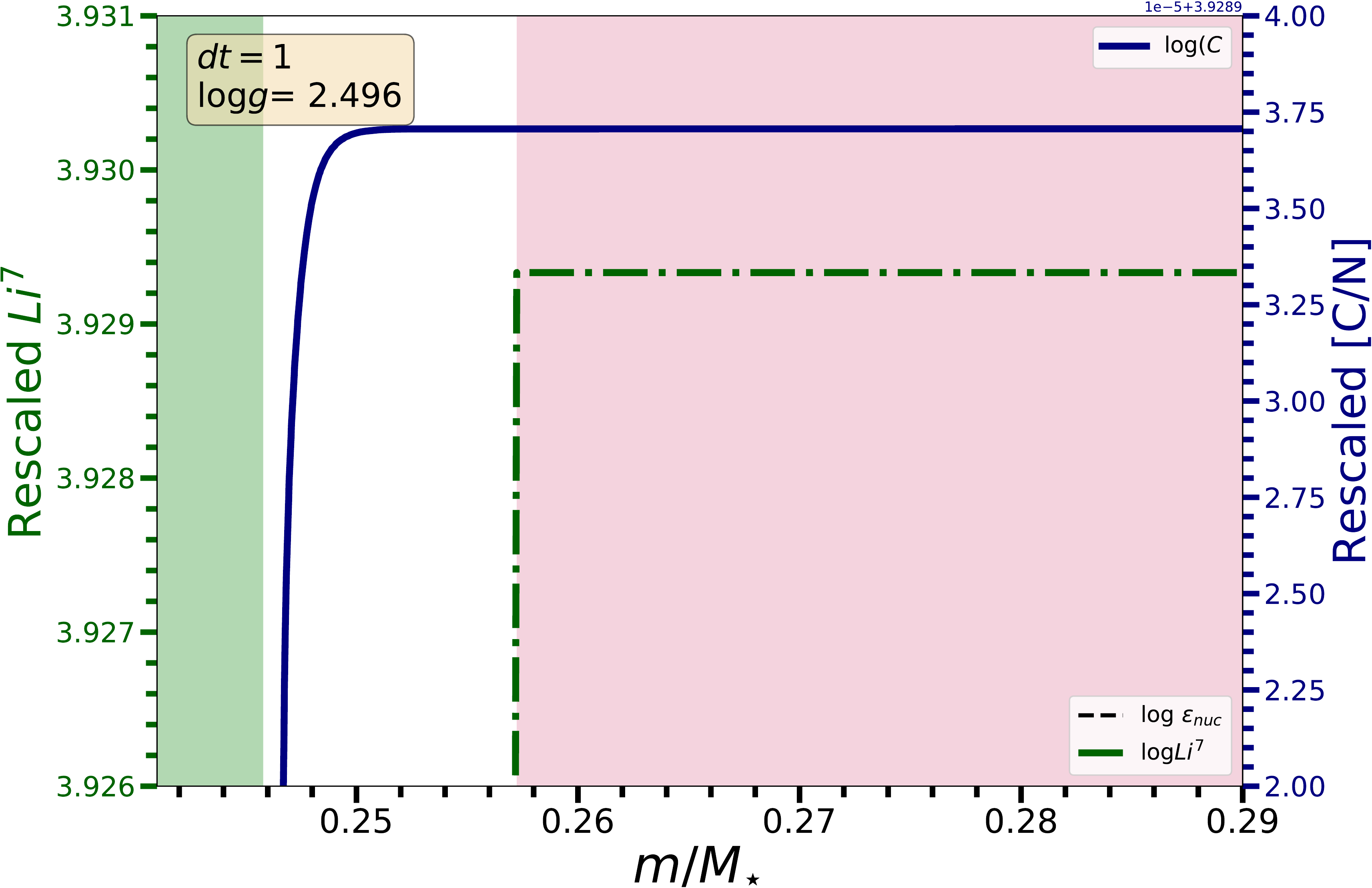}
\includegraphics[width=\columnwidth]{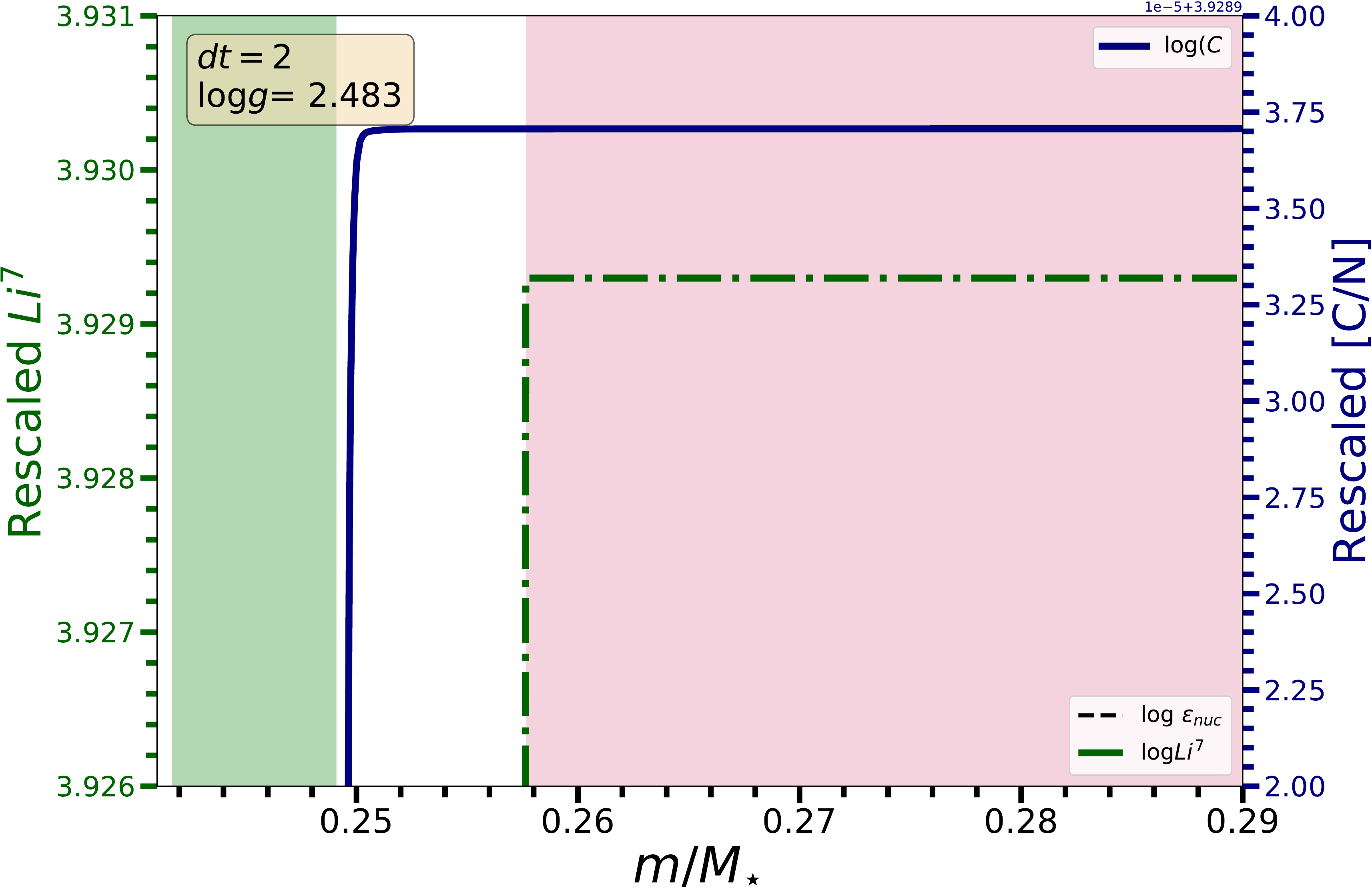}
\includegraphics[width=\columnwidth]{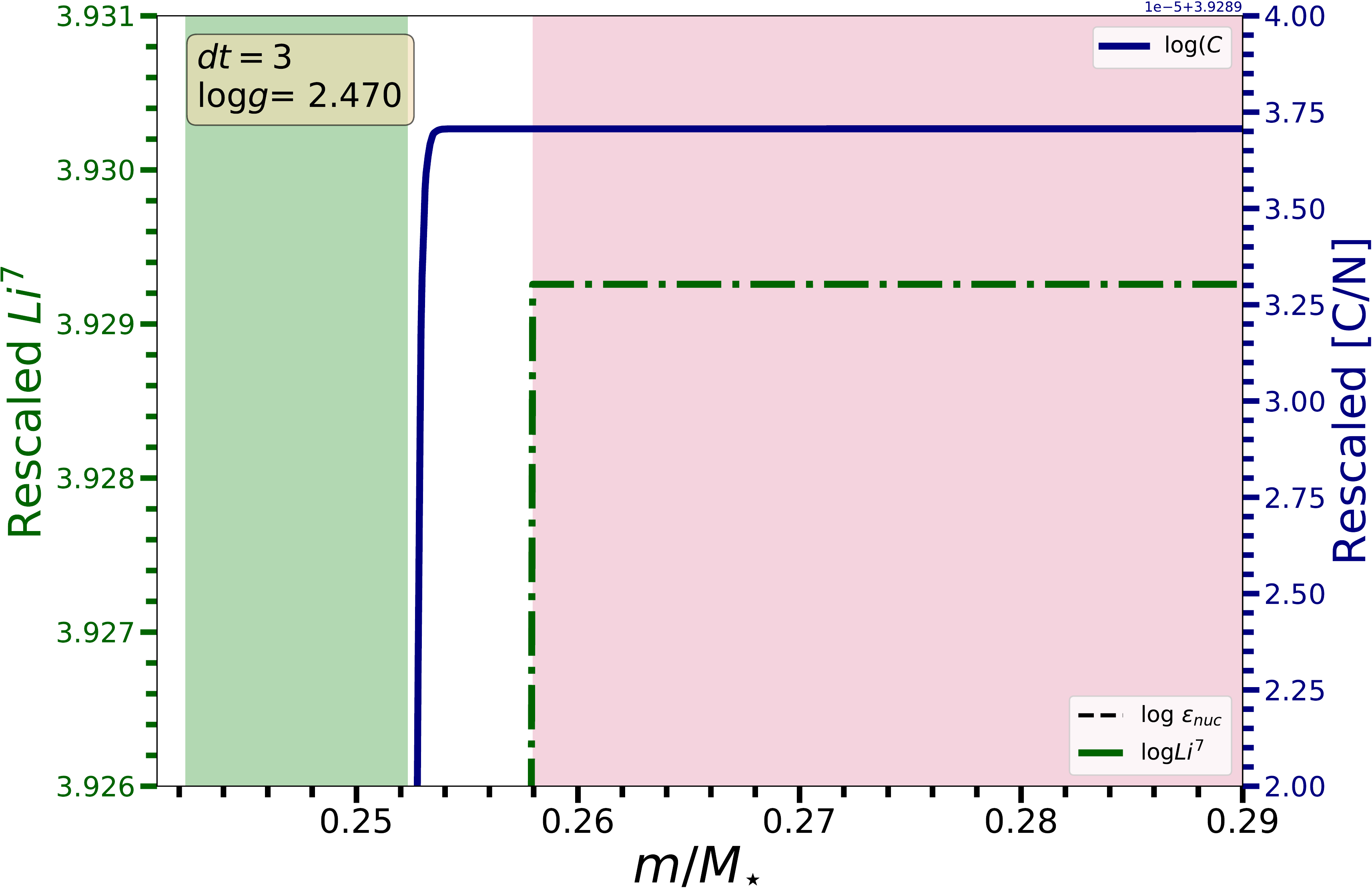}
\includegraphics[width=\columnwidth]{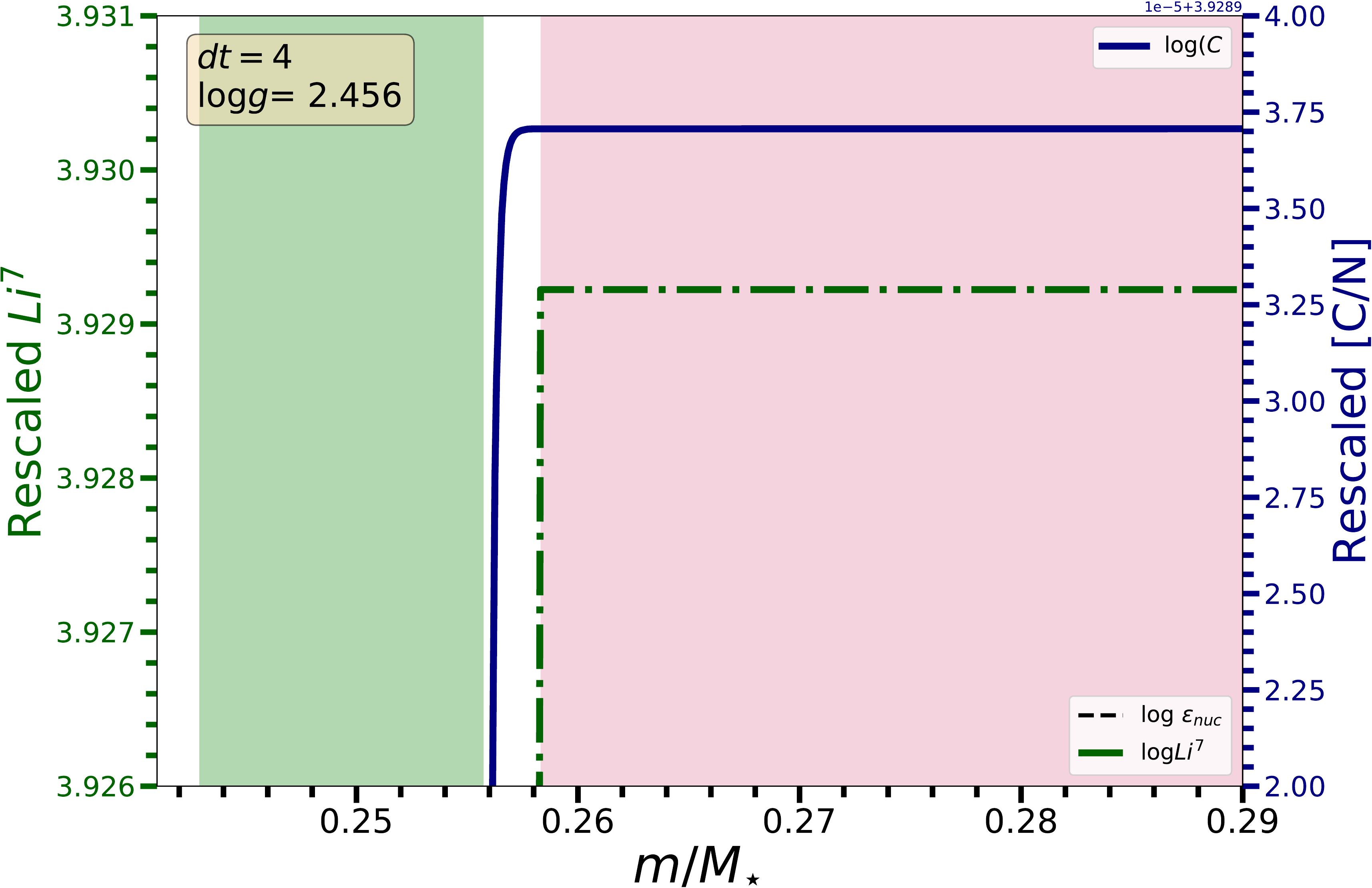}
\includegraphics[width=\columnwidth]{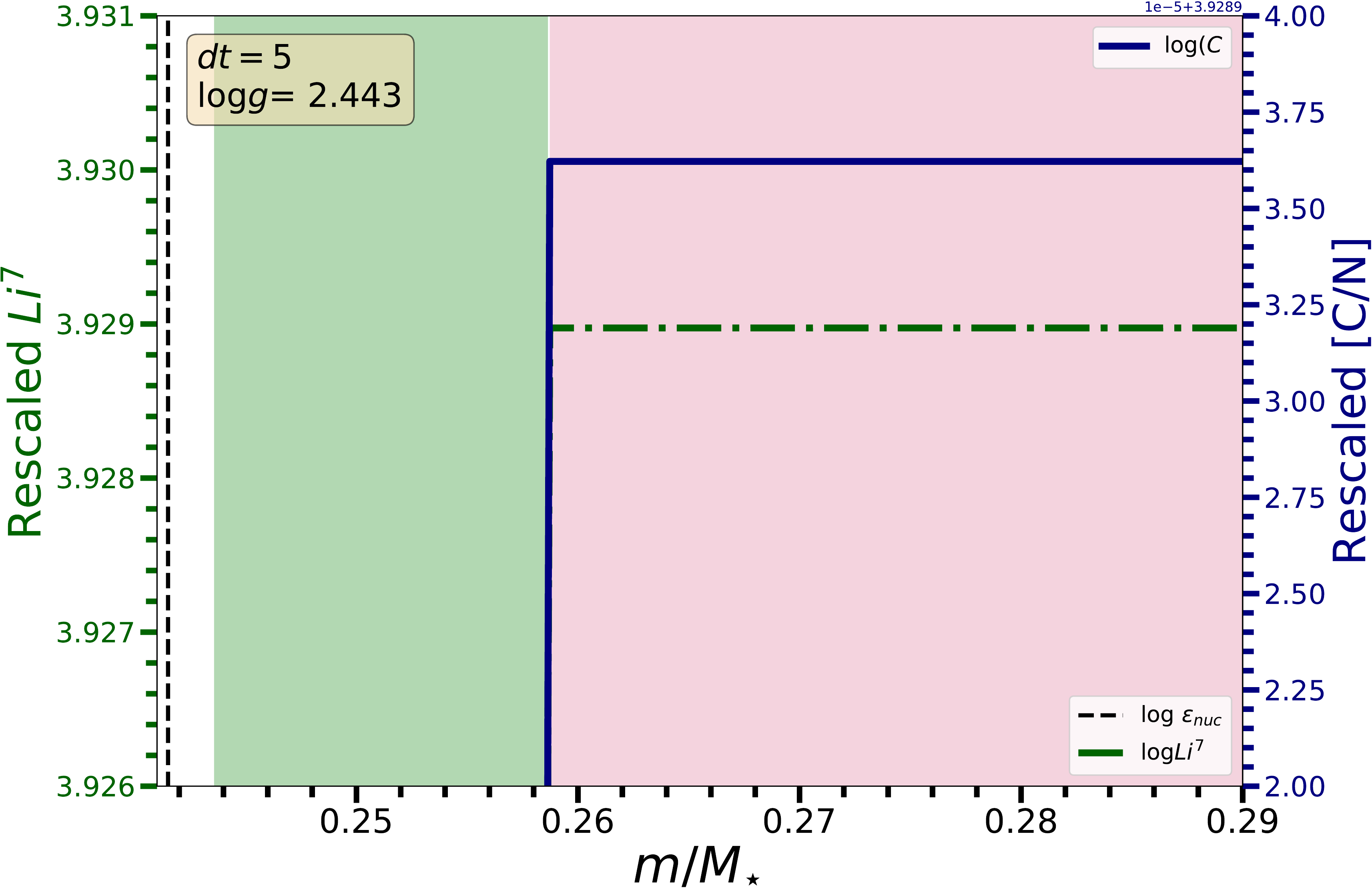}
\includegraphics[width=\columnwidth]{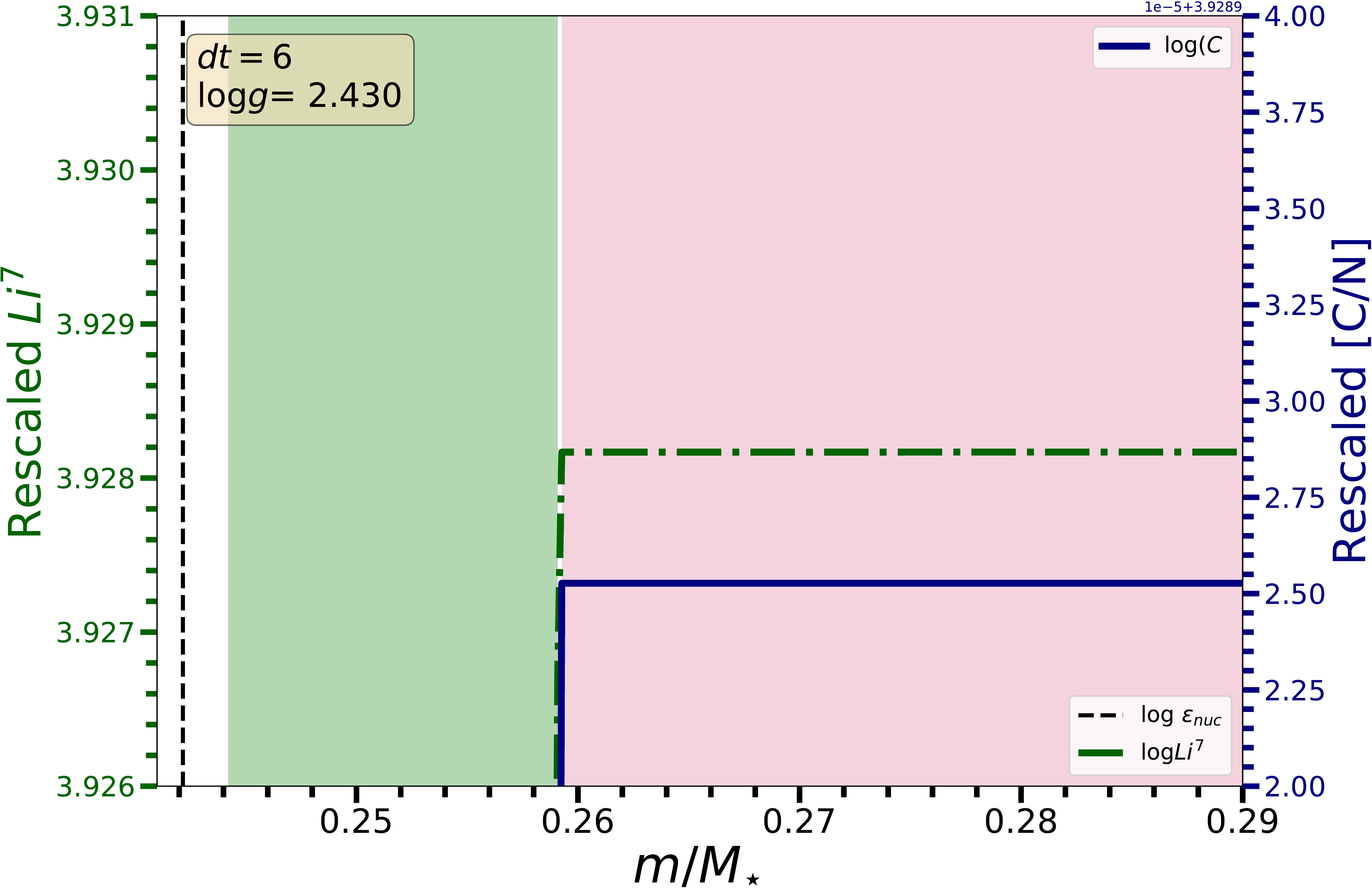}
\caption{Same as Figure \ref{fig:theory} but zooming in on the abundances to emphasize that the abundances of lithium (green dot-dashed line) and {(C/N)} (purple solid line) change only slightly until the thermohaline region (green) meets the base of the surface convection zone (pink), at which point the abundances of both elements at the surface begin to drop together. 
The abundances are rescaled so as to appear on similar axes by linearly shifting the $\log10$ abundance of lithium by a factor of roughly 15. We emphasize that we do not intend or attempt to provide predictions for specific values of abundances in this study: that is a much more difficult problem, impacted by numerous physical assumptions and their uncertainties that are not relevant to our investigation of the sequence of dilution events alone.
}
\label{fig:theory_abundances}
\end{center}
\end{figure*}

Our simulations trace the evolution of interior lithium, carbon, and nitrogen abundances and the nuclear energy generation front, $\varepsilon_{\text{nuc}}$, as well as the locations and extents of various mixing regions. Figure \ref{fig:theory} shows six snapshots of the interior near the base of the surface convection zone as a function of mass coordinate, $m/M_{\star}$, with corresponding evolutionary locations indicated by equivalent $dt$ on the HR diagram in Figure \ref{fig:HR}. The panels depict the surface-ward propagation and expansion of the thermohaline mixing front, shown in green, with increasing time: the region broadens and moves outward in mass coordinate from $dt=1$ to $dt=4$. At $dt=5$, the thermohaline region connects with the surface convection zone, shown in pink, which is receding towards the surface at a much slower rate over this time interval (the recession of the surface convection zone is apparent at enhanced scale, shown in the following figure). At $dt=6$, the regions have begun rapid mixing. Though panels $dt = 5$ and $dt = 6$ appear identical here, they represent summaries of the configuration before and after rapid mixing, respectively, as elaborated in Figure \ref{fig:theory_abundances}.

The black curve represents the nuclear burning shell, which also moves towards the surface over time. The blue and green curves show the interior abundance profiles of {(C$^{12}$/N$^{14}$)} and lithium, respectively, where the lithium abundance has been shifted uniformly so as to appear on the same vertical scale as the other quantities. The local minimum of the lithium abundance near $m/M_{\star}=0.25$ in panel $dt=1$ indicates the point at which the interior temperature is hot enough to destroy lithium. For detailed discussion on the behavior of the mean molecular weight profile and its inversion during the conditions that generate thermohaline mixing, see \citet{CharbonnelZahn2007} and \citet{CharbonnelLagarde2010}.

\begin{figure*}
\includegraphics[width=\columnwidth]{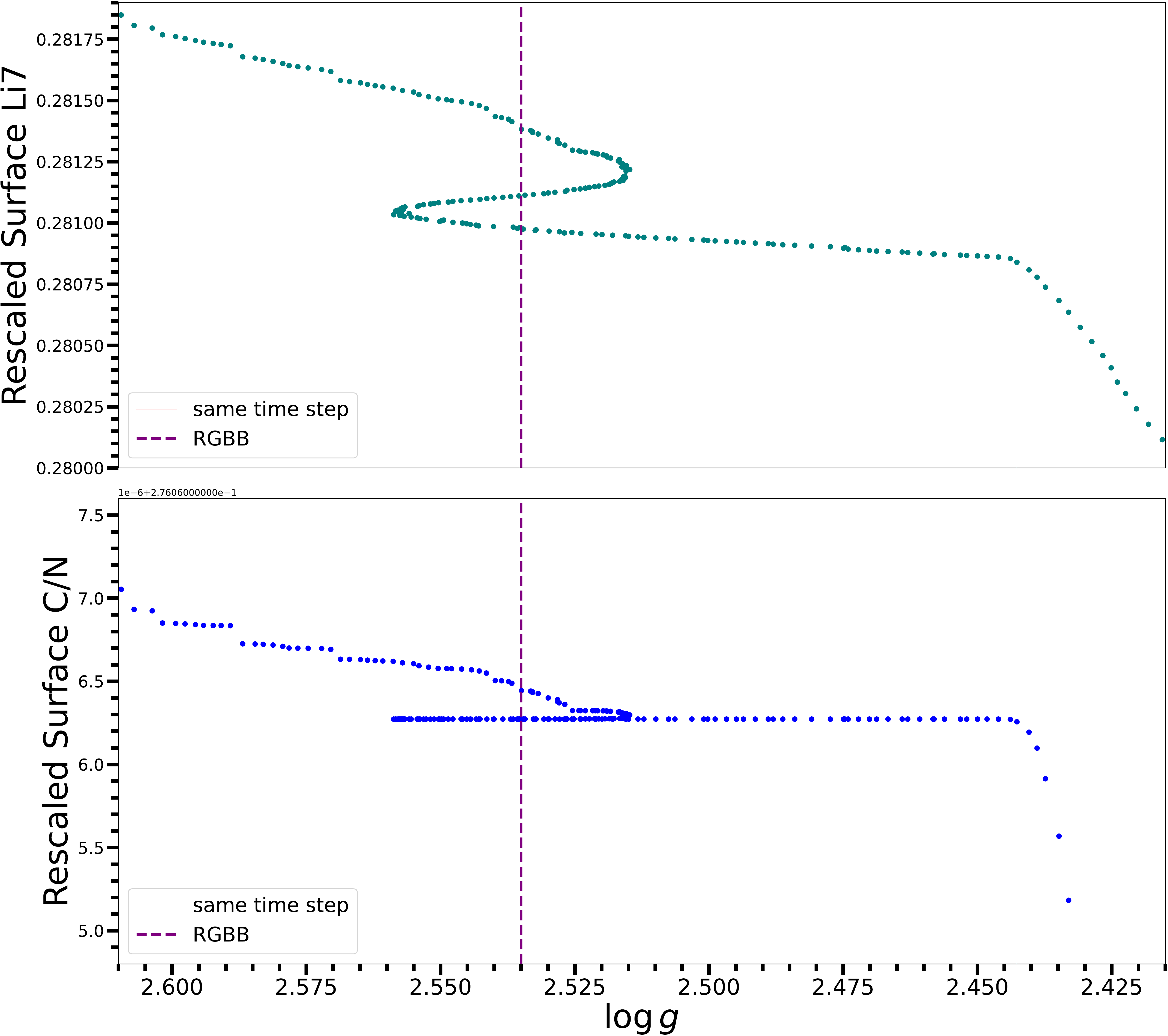}
\includegraphics[width=\columnwidth]{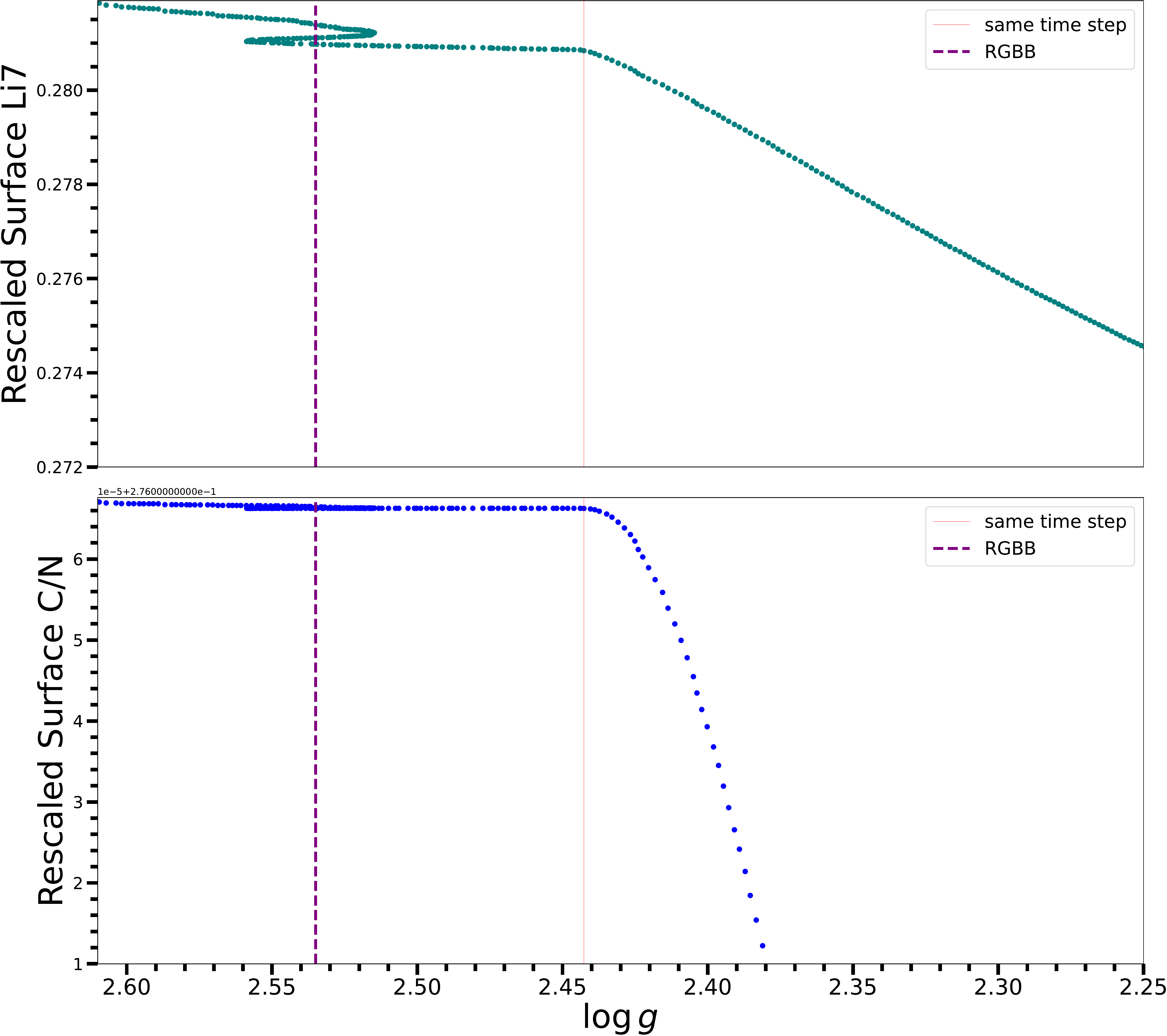}
\caption{Model predictions for the abundances of lithium (top) and [C/N] (bottom) are shown as the star evolves in surface gravity. The red vertical line marks the time step at which the dilution event related to the thermohaline front meeting the surface convection zone begins. This occurs slightly after the red giant branch bump, indicated with a purple dashed line. We show that this drop is predicted to begin at the same time step in both elements (emphasized in the zoomed-in panels on the left), and that the decline in the surface abundances at this point is significant compared to the slight changes visible earlier than and near the red giant branch bump (emphasized in the zoomed-out right panels). }
\label{fig:theorydrop}
\end{figure*}

Figure \ref{fig:theory_abundances} shows the same six structural snapshots, but rescaled so that the changes in abundance are visually detectable. As the panels show, there is a slow decrease in [C$^{12}$/N$^{14}$] and lithium as the time step increases from $dt=1$ to $dt=5$, but a sudden, much larger decrease in both at $dt=6$. In the final panel, we see that the drops in lithium and [C/N] occur at the same instant to within model precision (taken to be an uncertainty of half of one time step). This is the instant at which the thermohaline front meets the surface convection zone. We also observe this directly in Figure \ref{fig:theorydrop}, where we show the predicted abundances as a function of the surface gravity. In our models, the lithium and carbon-to-nitrogen ratios experience the onset of dilution at the same time step. The drop beginning at that point is significant relative to other changes in the abundances and should be visible in observational data. We argue that this conclusion is not sensitive to the numerical or physical choices made in the model. We emphasize, however, that absolute abundances and timings are extremely challenging to predict accurately given the wide variety of physical uncertainties in the models, \citep[see e.g.][Cinquegrana et al., \textit{subm.}, for more extensive discussion]{Lattanzio2015}. We are interested only in the relative changes here.

In general, this sequence is consistent with previous explorations of thermohaline mixing in low-mass red giant branch stars. After the RGBB, a region near the hydrogen burning shell becomes unstable to thermohaline mixing. On its own, thermohaline mixing in the model is insufficient to encourage mixing in the entire region between the burning shell and the base of the convection zone. However, when both rotational mixing and thermohaline mixing are included, the unstable region expands outwards, mixing through the region where lithium is destroyed, until it reaches the base of the surface convection zone. At this point, the star is able to (1) mix nitrogen enriched material from the shell up into the surface convection zone, and (2) to mix fragile elements like lithium from the cool surface of the star down into a region hot enough to destroy them. It is at this time that the surface abundances of lithium and [C/N] should change.

\section{Data}\label{sec:data}
The theoretical framework discussed above makes a clear prediction that the abundances of all elements affected by thermohaline mixing should begin to change at the same point. The recent explosion of available spectroscopic measurements of well characterized stars makes it possible to test this prediction empirically. 

For this analysis, we take the location of the red giant branch bump as a function of metallicity and the location of the change in the surface [C/N] ratio from \citet{Shetrone2019}.  The authors used a large dataset of $\alpha$-element enhanced stars from the SDSS-IV's \citep{Blanton2017} APOGEE-2 \citep{Majewski2017} Data Release 14 \citep{DR14}. These stars were observed by the APOGEE \citep{Wilson2019} spectrograph on the Sloan Foundation Telescope \citep{Gunn2006} for a variety of reasons \citep{Zasowski2017}, but processed and analyzed homogenously by the ASPCAP pipeline \citep{Nidever2015, Zamora2015, GarciaPerez2016, Holtzman2015, Holtzman2018}. {To separate red clump stars from the red giants used here, the authors used the spectroscopic evolutionary states estimated automatically from APOGEE \citep{Holtzman2015,Holtzman2018}, which use a combination of temperature, gravity, and luminosity information, and are calibrated based on the sample of APOGEE stars with asteroseismic estimates of evolutionary state \citep{Elsworth2017,Elsworth2019}.}

\begin{figure}
\begin{center}
\includegraphics[width=0.5\textwidth]{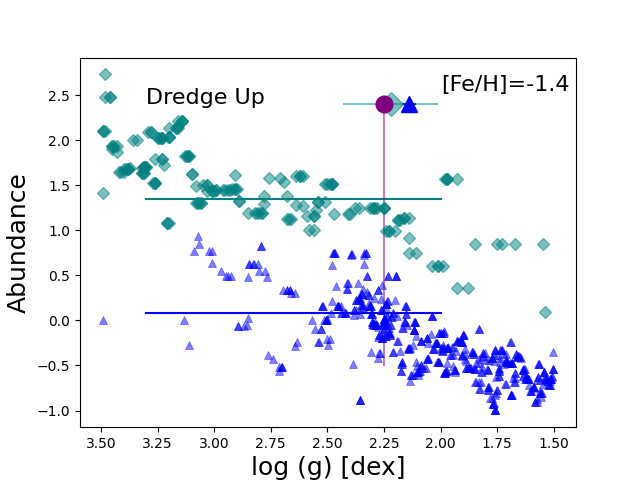}
\caption{With large samples of stars binned in metallicity, it is possible to trace chemical evolution as a function of surface gravity using lithium (teal diamonds) and [C/N] (blue triangles) as well as empirically identify the red giant branch bump (purple circle, and descending line for reference). In analogy with the figures in \citet{Shetrone2019} and in order to make changes in abundance more clear, we have indicated with lines the approximate level of the post-dredge-up, pre-bump abundances (teal and blue lines). }
\label{fig:dataraw}
\end{center}
\end{figure}

\begin{figure}
\begin{center}
\includegraphics[width=0.5\textwidth]{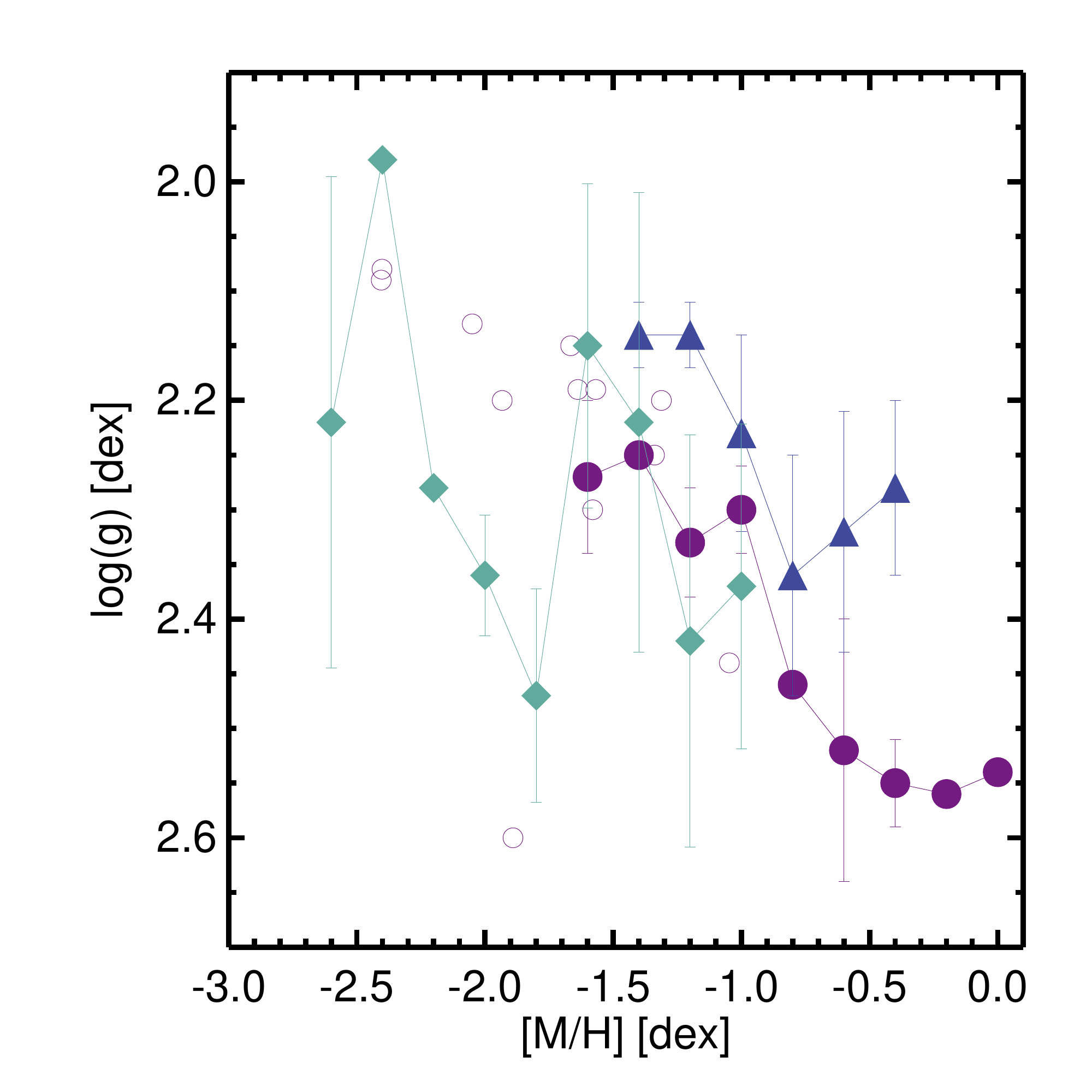}
\caption{The location in surface gravity, as a function of metallicity, of the drop in [C/N] associated with extra mixing (blue triangles), and the drop in lithium near the same point (teal diamonds). The empirical locations of the red giant branch bump from \citet{Shetrone2019} are marked in filled circles, and the estimated locations of the bump in \citet{Kirby2016} clusters that were studied by \citet{Nataf2013} are marked as open circles. While the drop in [C/N] is always above the red giant branch bump as expected, the drop in lithium abundance seems to happen significantly earlier, suggesting that a different mechanism is responsible.}
\label{fig:data}
\end{center}
\end{figure}

Since $\alpha$-enhanced stars were formed when the galaxy was young, these stars are assumed to be old, and therefore consistently low mass ($\sim 0.9$ \msun). \citet{Shetrone2019} therefore divided the full sample into metallicity bins, and studied each metallicity bin as if it represented an evolutionary sequence, roughly equivalent to a cluster. At each metallicity, they were able to identify the location of the overdensity of first ascent red giant branch stars that represented the red giant branch bump. In each bin, where possible, they also identified the location of the first dredge up, the extra mixing taking place near the red giant branch bump, and the location of any additional mixing on the upper giant branch if such mixing was detected at high significance (See Figure \ref{fig:dataraw}, or the original figures in that text for more detail). We show in Figure \ref{fig:data} their reported locations of the red giant branch bump and the extra mixing event, as probed by [C/N]. While the exact surface gravity of the red giant branch bump is metallicity dependent, the extra mixing event seems to happen consistently a few tenths of a dex above (after) the red giant branch bump. The decline of [C/N] at this point is consistent with the expectations from our theoretical understanding of thermohaline mixing as presented above. 

\begin{table}[htbp]
\caption{Locations on the giant branch for the drop in surface lithium abundance measured from the \citet{Kirby2016} data (see Section \ref{sec:data}) as well as the drop in [C/N] and the red giant branch bump as measured in \citet{Shetrone2019}. Error bars represent the uncertainty in the measured location from various methods rather than the systematic uncertainties on the data.}

\hspace{-0.5cm}\begin{tabular}{cccc}
 \hline \hline
{[M/H]} & {log(g)$_{\rm Li}$}& log(g)$_{\rm CN}$  & {log(g)$_{\rm bump}$}   \\ \hline
0.0 & ... & ...  &  2.54$\pm$0.01 \\ 
-0.2 & ...& ... &   2.56$\pm$0.01 \\ 
-0.4 & ... & {2.28}$\pm${0.08} & 2.55$\pm$0.04 \\ 
-0.6 & ... & {2.32}$\pm${0.11} & 2.52$\pm$0.12 \\ 
-0.8 & ...  & {2.36}$\pm${0.11} & 2.46$\pm$0.01 \\ 
-1.0 & 2.37$\pm$0.15 & {2.23}$\pm${0.09} & 2.30$\pm$0.04 \\ 
-1.2 & 2.42$\pm$0.19 & {2.14}$\pm${0.03} & 2.33$\pm$0.05 \\ 
-1.4 & 2.22$\pm$0.21 & {2.14} $\pm${0.03} & 2.25$\pm$0.01 \\ 
-1.6 & 2.15$\pm$0.15 &  ...  & 2.27$\pm$0.07 \\ 
-1.8 & 2.47$\pm$0.10 & ...   &  ... \\ 
-2.0 & 2.36$\pm$0.06 & ...   & ... \\ 
-2.2 & 2.28$\pm$0.01 & ...   & ... \\ 
-2.4 & 1.98$\pm$0.01 & ...   & ... \\ 
-2.6 & 2.22$\pm$0.22 & ...   & ... \\ 
\hline
\end{tabular}
\label{tab:thebins}
\end{table}

We supplement these data with measurements of the lithium abundance of thousands of red giants in globular clusters observed by \citet{Kirby2016}. From their dataset, we extract stars identified as first ascent red giants {based on the inspection of their colors and magnitudes relative to other stars in the cluster \citep[see][]{Kirby2016} for more discussion}. Following the procedure described in \citet{Shetrone2019}, we bin the stars in metallicity and search for a drop in lithium near the red giant branch bump (see data in Figure \ref{fig:dataraw}) by fitting a hyperbolic tangent function to the data between 1.5 and 2.9 dex in each metallicity bin . We report our results in Table \ref{tab:thebins}. Unfortunately, there are insufficient data points in this sample to robustly detect the red giant branch bump location directly  in the \citet{Kirby2016} spectroscopic sample. However some, of these clusters were studied with HST by \citet{Nataf2013}, and the V magnitude of their red giant branch bumps were measured. For those clusters, we used stars near that V magnitude with measured surface gravities to estimate the surface gravity of the red giant branch bump on the \citet{Kirby2016} scale, and used the average metallicity measured by \citet{Kirby2016} for the whole cluster. While there may be uncertainties and systematics from this method for individual clusters, the ensemble seems to match well the trend in bump surface gravity measured in \citet{Shetrone2019} (Figure \ref{fig:data}). We therefore make the assumption that the two measurements are on the same scale and can be compared, a point to which we return in the discussion section.

We show in Figure \ref{fig:data} that a decrease in lithium abundance does indeed happen near the red giant branch bump, but it is not coincident with the drop in [C/N]. This implies that the two changes cannot be due to the same thermohaline mixing event.

\section{Discussion}
In this paper, we have presented an argument regarding the the observed timing of chemical depletion events and the ability of thermohaline mixing alone to explain them. Our simulations clearly show it cannot, thus inviting a more sophisticated theoretical explanation for the observed time-separation of abundance depletion trends.

Given the observational data available, the extra mixing events that alter the surface abundances of [C/N] and lithium cannot both be ascribed to thermohaline mixing since they do not occur at the same time. The regions that must be mixed have previously been homogenized during the first dredge up, and therefore should not retain chemical gradients and discontinuities left behind from pre-main-sequence or main sequence processes. As shown in the literature and by the modeling above, the thermohaline mixing front is propagating outwards from the core after the bump, and therefore extra mixing events related to thermohaline mixing should not begin to alter the surface chemistry until about 0.1 dex above the red giant branch bump. While this result is consistent with the measured location of the change in [C/N], it is inconsistent with the location of the change of the lithium abundance. {Such discrepancies have been noted in the past in detailed cluster analyses  \citep{Angelou2015}, but in general it has been argued that they result from other other issues in the analysis such as the choice of distance modulus \citep{Henkel2017}. Under the assumption that the offset we see in our data is real, we}  first discuss possible theoretical explanations for this result, and then suggest several observational tests which could ensure that it is robust.

Theoretically, in order to obtain a drop in lithium before the thermohaline front has met the base of the convection zone, there needs to be sufficient extra mixing beneath the convective envelope to penetrate into the region of the star that is hot enough to destroy lithium. Simulations of convective zones do often see convective-like flows moving into regions that are formally stably stratified, variously called convective overshoot \citep[e.g.][]{Zahn1991, Korre2019}, entrainment \citep[e.g.][]{MeakinArnett2007, Horst2021} or convective penetration \citep [e.g.][]{Hurlburt1994, Anders2021}. However, both simulations and observations \citep{ClaretTorres2019, Pedersen2021} of main sequence and red giant branch stars generally find these overshoot regions to be significantly smaller than a pressure scale height \citep{Joyce2018a}. In the case of red giant stars, the lithium burning zone is more than a pressure scale height below the formal boundary of the surface convection zone \citep{Straus1976}. Rotational mixing can be present in this region in models \citep[see e.g.][]{Lagarde2012, Charbonnel2020}, but it is generally not expected to be strong enough to deplete a significant amount of lithium on its own. This suggests that some unusual process, or combination of processes, may be acting in this region. 

However, we also acknowledge the possibility that the offset between the lithium and [C/N] mixing locations could be the result of observational biases. In particular, calibrating the absolute surface gravities and metallicities between surveys can be challenging, and so exploration of a sample of stars where both lithium and [C/N] have been measured would be a strong test of this result, as would detailed analysis of both elements in a single cluster, where systematics are easier to control and differential measurements are extremely powerful. Larger sample sizes, from, e.g., the more recent APOGEE DR17 data or GALAH DR3, would also reduce the observational uncertainty in the location of the drop in abundance for each element, as well as the location and size of the luminosity bump. More extensive catalogs might also be better able to map the shape of the change near the dilution drop. They may also be better able to account for the expected differences in the rate of change of lithium versus the carbon-to-nitrogen ratio, which could affect the inferred location of the onset of the mixing. Additional mixing diagnostics such as \ctwelvecthirteen\ might also be illuminating, as would a study of the mass dependence of the location and depth of extra mixing. Mis-identification of evolutionary states in the data could also alter the measured location of the extra mixing: for example, dragging the estimated drop from the bump towards the clump, where stars are already mixed. This would present as a shift towards higher gravity at low metallicity and towards a lower gravity at high metallicity. While we believe that this is not a significant factor for our data sets, where evolutionary states have been carefully calibrated or curated by eye, it is important to consider for future investigations of this type.

This investigation serves as an exciting foray into the new reality where the properties of extremely large samples of stars are exquisitely characterized, and therefore even minute surface changes can be identified observationally and used to test theories of stellar interiors and evolution. As spectroscopic surveys release even more data (e.g. GALAH, Gaia-ESO, SDSS-V), and as these data can be combined with other methods of characterizing stars, including astrometry, asteroseismology, interferometry, photometry, and time domain studies, we venture into a regime where a theory must not only qualitatively match the expected results, but also quantitatively agree with precisely characterized observational data, with significantly reduced systematic uncertainties.

\section*{Acknowledgements}
J.T. and M.J. contributed equally to this manuscript. The authors gratefully acknowledge helpful discussions with Adrian Fraser, Matteo Cantiello, Evan Anders, and the rest of the thermohaline working group at the KITP Probes of Transport in Stars 2021 meeting. J.T. would like to thank Evan Kirby for discussions that contributed to the beginnings of this work and Ann Boesgaard for pointing out useful references. 
M.J. wishes to thank Brian Chaboyer for useful discussion on numerics during the second revision of this manuscript. J.T. and M.J. also wish to thank Amanda Karakas, Marc Pinsonneault, and L\'aszl\'o Moln\'ar, each of whom provided an independent reading and review of this manuscript. We thank the referees for their comments that improved this manuscript.

J.T. acknowledges that support for this work was provided by NASA through the NASA Hubble Fellowship grant No.51424 awarded by the Space Telescope Science Institute, which is operated by the Association of Universities for Research in Astronomy, Inc., for NASA, under contract NAS5-26555.
M.J.\ was supported by the Lasker Data Science Fellowship, awarded by the Space Telescope Science Institute, and the Kavli Institute for Theoretical Physics at the University of California, Santa Barbara.
This research was supported in part by the National Science Foundation under Grant No. NSF PHY-1748958.

\appendix
\section{MESA Details}
\label{appen:mesa}

The MESA EOS is a blend of the OPAL \citep{Rogers2002}, SCVH
\citep{Saumon1995}, FreeEOS \citep{Irwin2004}, HELM \citep{Timmes2000},
and PC \citep{Potekhin2010} EOSes.

Radiative opacities are primarily from OPAL \citep{Iglesias1993,
Iglesias1996}, with low-temperature data from \citet{Ferguson2005}
and the high-temperature, Compton-scattering dominated regime by
\citet{Buchler1976}.  Electron conduction opacities are from
\citet{Cassisi2007}.

Nuclear reaction rates are from JINA REACLIB \citep{Cyburt2010} plus
additional tabulated weak reaction rates \citet{Fuller1985, Oda1994, Langanke2000}.
Screening is included via the prescription of \citet{Chugunov2007}.
Thermal neutrino loss rates are from \citet{Itoh1996}.

\bibliographystyle{apj}
\bibliography{thermohaline.bib, library.bib, library2.bib, ms.bib, msCN.bib, mesa.bib, Joyce.bib} 

\begin{thebibliography}{}
\expandafter\ifx\csname natexlab\endcsname\relax\def\natexlab#1{#1}\fi

\bibitem[{{Abolfathi} {et~al.}(2018){Abolfathi}, {Aguado}, {Aguilar}, {Allende
  Prieto}, {Almeida}, {Ananna}, {Anders}, {Anderson}, {Andrews}, {Anguiano}, \&
  et~al.}]{DR14}
{Abolfathi}, B., {Aguado}, D.~S., {Aguilar}, G., {et~al.} 2018, \apjs, 235, 42

\bibitem[{{Anders} {et~al.}(2021){Anders}, {Jermyn}, {Lecoanet}, \&
  {Brown}}]{Anders2021}
{Anders}, E.~H., {Jermyn}, A.~S., {Lecoanet}, D., \& {Brown}, B.~P. 2021, arXiv
  e-prints, arXiv:2110.11356

\bibitem[{{Anders} {et~al.}(2022){Anders}, {Jermyn}, {Lecoanet}, {Fraser},
  {Cresswell}, {Joyce}, \& {Fuentes}}]{Anders2022}
{Anders}, E.~H., {Jermyn}, A.~S., {Lecoanet}, D., {et~al.} 2022, arXiv
  e-prints, arXiv:2203.06186

\bibitem[{{Angelou} {et~al.}(2015){Angelou}, {D'Orazi}, {Constantino},
  {Church}, {Stancliffe}, \& {Lattanzio}}]{Angelou2015}
{Angelou}, G.~C., {D'Orazi}, V., {Constantino}, T.~N., {et~al.} 2015, \mnras,
  450, 2423

\bibitem[{{Blanton} {et~al.}(2017){Blanton}, {Bershady}, {Abolfathi},
  {Albareti}, {Allende Prieto}, {Almeida}, {Alonso-Garc{\'{\i}}a}, {Anders},
  {Anderson}, {Andrews}, \& et~al.}]{Blanton2017}
{Blanton}, M.~R., {Bershady}, M.~A., {Abolfathi}, B., {et~al.} 2017, \aj, 154,
  28

\bibitem[{{Boesgaard} {et~al.}(2020){Boesgaard}, {Lum}, \&
  {Deliyannis}}]{Boesgaard2020}
{Boesgaard}, A.~M., {Lum}, M.~G., \& {Deliyannis}, C.~P. 2020, \apj, 888, 28

\bibitem[{{B{\"o}hm-Vitense}(1958)}]{BoehmVitense1958}
{B{\"o}hm-Vitense}, E. 1958, \zap, 46, 108

\bibitem[{{Boothroyd} \& {Sackmann}(1999)}]{BoothroydSackmann1999}
{Boothroyd}, A.~I., \& {Sackmann}, I.~J. 1999, \apj, 510, 232

\bibitem[{{Brown} {et~al.}(2013){Brown}, {Garaud}, \& {Stellmach}}]{Brown2013}
{Brown}, J.~M., {Garaud}, P., \& {Stellmach}, S. 2013, \apj, 768, 34

\bibitem[{{Buchler} \& {Yueh}(1976)}]{Buchler1976}
{Buchler}, J.~R., \& {Yueh}, W.~R. 1976, \apj, 210, 440

\bibitem[{{Cantiello} \& {Langer}(2010)}]{CantielloLanger2010}
{Cantiello}, M., \& {Langer}, N. 2010, \aap, 521, A9

\bibitem[{{Carbon} {et~al.}(1982){Carbon}, {Romanishin}, {Langer}, {Butler},
  {Kemper}, {Trefzger}, {Kraft}, \& {Suntzeff}}]{Carbon1982}
{Carbon}, D.~F., {Romanishin}, W., {Langer}, G.~E., {et~al.} 1982, \apjs, 49,
  207

\bibitem[{{Cassisi} {et~al.}(2007){Cassisi}, {Potekhin}, {Pietrinferni},
  {Catelan}, \& {Salaris}}]{Cassisi2007}
{Cassisi}, S., {Potekhin}, A.~Y., {Pietrinferni}, A., {Catelan}, M., \&
  {Salaris}, M. 2007, \apj, 661, 1094

\bibitem[{{Chanam{\'e}} {et~al.}(2005){Chanam{\'e}}, {Pinsonneault}, \&
  {Terndrup}}]{Chaname2005}
{Chanam{\'e}}, J., {Pinsonneault}, M., \& {Terndrup}, D.~M. 2005, \apj, 631,
  540

\bibitem[{{Charbonnel} \& {Lagarde}(2010)}]{CharbonnelLagarde2010}
{Charbonnel}, C., \& {Lagarde}, N. 2010, \aap, 522, A10

\bibitem[{{Charbonnel} {et~al.}(1994){Charbonnel}, {Vauclair}, {Maeder},
  {Meynet}, \& {Schaller}}]{Charbonnel1994}
{Charbonnel}, C., {Vauclair}, S., {Maeder}, A., {Meynet}, G., \& {Schaller}, G.
  1994, \aap, 283, 155

\bibitem[{{Charbonnel} \& {Zahn}(2007{\natexlab{a}})}]{CharbonnelZahn2007b}
{Charbonnel}, C., \& {Zahn}, J.~P. 2007{\natexlab{a}}, \aap, 476, L29

\bibitem[{{Charbonnel} \& {Zahn}(2007{\natexlab{b}})}]{CharbonnelZahn2007}
---. 2007{\natexlab{b}}, \aap, 467, L15

\bibitem[{{Charbonnel} {et~al.}(2020){Charbonnel}, {Lagarde}, {Jasniewicz},
  {North}, {Shetrone}, {Krugler Hollek}, {Smith}, {Smiljanic}, {Palacios}, \&
  {Ottoni}}]{Charbonnel2020}
{Charbonnel}, C., {Lagarde}, N., {Jasniewicz}, G., {et~al.} 2020, \aap, 633,
  A34

\bibitem[{{Chugunov} {et~al.}(2007){Chugunov}, {Dewitt}, \&
  {Yakovlev}}]{Chugunov2007}
{Chugunov}, A.~I., {Dewitt}, H.~E., \& {Yakovlev}, D.~G. 2007, \prd, 76, 025028

\bibitem[{{Claret} \& {Torres}(2019)}]{ClaretTorres2019}
{Claret}, A., \& {Torres}, G. 2019, \apj, 876, 134

\bibitem[{{Cox}(1980)}]{Cox1980}
{Cox}, J.~P. 1980, {Theory of stellar pulsation}

\bibitem[{{Cox} \& {Giuli}(1968)}]{CoxGiuli1968}
{Cox}, J.~P., \& {Giuli}, R.~T. 1968, {Principles of stellar structure}

\bibitem[{{Cyburt} {et~al.}(2010){Cyburt}, {Amthor}, {Ferguson}, {Meisel},
  {Smith}, {Warren}, {Heger}, {Hoffman}, {Rauscher}, {Sakharuk}, {Schatz},
  {Thielemann}, \& {Wiescher}}]{Cyburt2010}
{Cyburt}, R.~H., {Amthor}, A.~M., {Ferguson}, R., {et~al.} 2010, \apjs, 189,
  240

\bibitem[{{Denissenkov}(2010)}]{Denissenkov2010}
{Denissenkov}, P.~A. 2010, \apj, 723, 563

\bibitem[{{Denissenkov} \& {Merryfield}(2011)}]{Denissenkov2011}
{Denissenkov}, P.~A., \& {Merryfield}, W.~J. 2011, \apjl, 727, L8

\bibitem[{{Eddington}(1916)}]{Eddington1916}
{Eddington}, A.~S. 1916, \mnras, 77, 16

\bibitem[{{Eggleton} {et~al.}(2006){Eggleton}, {Dearborn}, \&
  {Lattanzio}}]{Eggleton2006}
{Eggleton}, P.~P., {Dearborn}, D. S.~P., \& {Lattanzio}, J.~C. 2006, Science,
  314, 1580

\bibitem[{{Elsworth} {et~al.}(2017){Elsworth}, {Hekker}, {Basu}, \&
  {Davies}}]{Elsworth2017}
{Elsworth}, Y., {Hekker}, S., {Basu}, S., \& {Davies}, G.~R. 2017, \mnras, 466,
  3344

\bibitem[{{Elsworth} {et~al.}(2019){Elsworth}, {Hekker}, {Johnson},
  {Kallinger}, {Mosser}, {Pinsonneault}, {Hon}, {Kuszlewicz}, {Miglio},
  {Serenelli}, {Stello}, {Tayar}, \& {Vrard}}]{Elsworth2019}
{Elsworth}, Y., {Hekker}, S., {Johnson}, J.~A., {et~al.} 2019, \mnras, 489,
  4641

\bibitem[{{Ferguson} {et~al.}(2005){Ferguson}, {Alexander}, {Allard}, {Barman},
  {Bodnarik}, {Hauschildt}, {Heffner-Wong}, \& {Tamanai}}]{Ferguson2005}
{Ferguson}, J.~W., {Alexander}, D.~R., {Allard}, F., {et~al.} 2005, \apj, 623,
  585

\bibitem[{{Fraser} {et~al.}(2022){Fraser}, {Joyce}, {Anders}, {Tayar}, \&
  {Cantiello}}]{Fraser2022}
{Fraser}, A.~E., {Joyce}, M., {Anders}, E.~H., {Tayar}, J., \& {Cantiello}, M.
  2022, arXiv e-prints, arXiv:2204.08487

\bibitem[{{Fuller} {et~al.}(1985){Fuller}, {Fowler}, \& {Newman}}]{Fuller1985}
{Fuller}, G.~M., {Fowler}, W.~A., \& {Newman}, M.~J. 1985, \apj, 293, 1

\bibitem[{{Fusi Pecci} {et~al.}(1990){Fusi Pecci}, {Ferraro}, {Crocker},
  {Rood}, \& {Buonanno}}]{FusiPecci1990}
{Fusi Pecci}, F., {Ferraro}, F.~R., {Crocker}, D.~A., {Rood}, R.~T., \&
  {Buonanno}, R. 1990, \aap, 238, 95

\bibitem[{{Garaud} {et~al.}(2019){Garaud}, {Kumar}, \& {Sridhar}}]{Garaud2019}
{Garaud}, P., {Kumar}, A., \& {Sridhar}, J. 2019, \apj, 879, 60

\bibitem[{{Garc{\'{\i}}a P{\'e}rez} {et~al.}(2016){Garc{\'{\i}}a P{\'e}rez},
  {Allende Prieto}, {Holtzman}, {Shetrone}, {M{\'e}sz{\'a}ros}, {Bizyaev},
  {Carrera}, {Cunha}, {Garc{\'{\i}}a-Hern{\'a}ndez}, {Johnson}, {Majewski},
  {Nidever}, {Schiavon}, {Shane}, {Smith}, {Sobeck}, {Troup}, {Zamora},
  {Weinberg}, {Bovy}, {Eisenstein}, {Feuillet}, {Frinchaboy}, {Hayden},
  {Hearty}, {Nguyen}, {O'Connell}, {Pinsonneault}, {Wilson}, \&
  {Zasowski}}]{GarciaPerez2016}
{Garc{\'{\i}}a P{\'e}rez}, A.~E., {Allende Prieto}, C., {Holtzman}, J.~A.,
  {et~al.} 2016, \aj, 151, 144

\bibitem[{{Gratton} {et~al.}(2000){Gratton}, {Sneden}, {Carretta}, \&
  {Bragaglia}}]{Gratton2000}
{Gratton}, R.~G., {Sneden}, C., {Carretta}, E., \& {Bragaglia}, A. 2000, \aap,
  354, 169

\bibitem[{{Grevesse} \& {Sauval}(1998)}]{GrevesseSauval1998}
{Grevesse}, N., \& {Sauval}, A.~J. 1998, \ssr, 85, 161

\bibitem[{{Gunn} {et~al.}(2006){Gunn}, {Siegmund}, {Mannery}, {Owen}, {Hull},
  {Leger}, {Carey}, {Knapp}, {York}, {Boroski}, {Kent}, {Lupton}, {Rockosi},
  {Evans}, {Waddell}, {Anderson}, {Annis}, {Barentine}, {Bartoszek}, {Bastian},
  {Bracker}, {Brewington}, {Briegel}, {Brinkmann}, {Brown}, {Carr},
  {Czarapata}, {Drennan}, {Dombeck}, {Federwitz}, {Gillespie}, {Gonzales},
  {Hansen}, {Harvanek}, {Hayes}, {Jordan}, {Kinney}, {Klaene}, {Kleinman},
  {Kron}, {Kresinski}, {Lee}, {Limmongkol}, {Lindenmeyer}, {Long}, {Loomis},
  {McGehee}, {Mantsch}, {Neilsen}, {Neswold}, {Newman}, {Nitta}, {Peoples},
  {Pier}, {Prieto}, {Prosapio}, {Rivetta}, {Schneider}, {Snedden}, \&
  {Wang}}]{Gunn2006}
{Gunn}, J.~E., {Siegmund}, W.~A., {Mannery}, E.~J., {et~al.} 2006, \aj, 131,
  2332

\bibitem[{{Henkel} {et~al.}(2018){Henkel}, {Karakas}, {Casey}, {Church}, \&
  {Lattanzio}}]{Henkel2018}
{Henkel}, K., {Karakas}, A.~I., {Casey}, A.~R., {Church}, R.~P., \&
  {Lattanzio}, J.~C. 2018, \apjl, 863, L5

\bibitem[{{Henkel} {et~al.}(2017){Henkel}, {Karakas}, \&
  {Lattanzio}}]{Henkel2017}
{Henkel}, K., {Karakas}, A.~I., \& {Lattanzio}, J.~C. 2017, \mnras, 469, 4600

\bibitem[{{Holtzman} {et~al.}(2015){Holtzman}, {Shetrone}, {Johnson}, {Allende
  Prieto}, {Anders}, {Andrews}, {Beers}, {Bizyaev}, {Blanton}, {Bovy},
  {Carrera}, {Chojnowski}, {Cunha}, {Eisenstein}, {Feuillet}, {Frinchaboy},
  {Galbraith-Frew}, {Garc{\'{\i}}a P{\'e}rez}, {Garc{\'{\i}}a-Hern{\'a}ndez},
  {Hasselquist}, {Hayden}, {Hearty}, {Ivans}, {Majewski}, {Martell},
  {Meszaros}, {Muna}, {Nidever}, {Nguyen}, {O'Connell}, {Pan}, {Pinsonneault},
  {Robin}, {Schiavon}, {Shane}, {Sobeck}, {Smith}, {Troup}, {Weinberg},
  {Wilson}, {Wood-Vasey}, {Zamora}, \& {Zasowski}}]{Holtzman2015}
{Holtzman}, J.~A., {Shetrone}, M., {Johnson}, J.~A., {et~al.} 2015, \aj, 150,
  148

\bibitem[{{Holtzman} {et~al.}(2018){Holtzman}, {Hasselquist}, {Shetrone},
  {Cunha}, {Allende Prieto}, {Anguiano}, {Bizyaev}, {Bovy}, {Casey},
  {Edvardsson}, {Johnson}, {J{\"o}nsson}, {Meszaros}, {Smith}, {Sobeck},
  {Zamora}, {Chojnowski}, {Fernandez-Trincado}, {Garcia-Hernandez}, {Majewski},
  {Pinsonneault}, {Souto}, {Stringfellow}, {Tayar}, {Troup}, \&
  {Zasowski}}]{Holtzman2018}
{Holtzman}, J.~A., {Hasselquist}, S., {Shetrone}, M., {et~al.} 2018, \aj, 156,
  125

\bibitem[{{Horst} {et~al.}(2021){Horst}, {Hirschi}, {Edelmann}, {Andr{\'a}ssy},
  \& {R{\"o}pke}}]{Horst2021}
{Horst}, L., {Hirschi}, R., {Edelmann}, P.~V.~F., {Andr{\'a}ssy}, R., \&
  {R{\"o}pke}, F.~K. 2021, \aap, 653, A55

\bibitem[{{Hurlburt} {et~al.}(1994){Hurlburt}, {Toomre}, {Massaguer}, \&
  {Zahn}}]{Hurlburt1994}
{Hurlburt}, N.~E., {Toomre}, J., {Massaguer}, J.~M., \& {Zahn}, J.-P. 1994,
  \apj, 421, 245

\bibitem[{{Iben}(1964)}]{iben1964}
{Iben}, Jr., I. 1964, \apj, 140, 1631

\bibitem[{{Iben}(1967)}]{Iben1967}
---. 1967, \apj, 147, 624

\bibitem[{{Iglesias} \& {Rogers}(1993)}]{Iglesias1993}
{Iglesias}, C.~A., \& {Rogers}, F.~J. 1993, \apj, 412, 752

\bibitem[{{Iglesias} \& {Rogers}(1996{\natexlab{a}})}]{IglesiasRogers1996}
---. 1996{\natexlab{a}}, \apj, 464, 943

\bibitem[{{Iglesias} \& {Rogers}(1996{\natexlab{b}})}]{Iglesias1996}
---. 1996{\natexlab{b}}, \apj, 464, 943

\bibitem[{{Irwin}(2004)}]{Irwin2004}
{Irwin}, A.~W. 2004, {The FreeEOS Code for Calculating the Equation of State
  for Stellar Interiors I: An Improved EFF-Style Approximation for the
  Fermi-Dirac Integrals}

\bibitem[{{Itoh} {et~al.}(1996){Itoh}, {Hayashi}, {Nishikawa}, \&
  {Kohyama}}]{Itoh1996}
{Itoh}, N., {Hayashi}, H., {Nishikawa}, A., \& {Kohyama}, Y. 1996, \apjs, 102,
  411

\bibitem[{{Joyce} \& {Chaboyer}(2015)}]{Joyce2015}
{Joyce}, M., \& {Chaboyer}, B. 2015, \apj, 814, 142

\bibitem[{{Joyce} \& {Chaboyer}(2018)}]{Joyce2018a}
---. 2018, \apj, 856, 10

\bibitem[{{Khan} {et~al.}(2018){Khan}, {Hall}, {Miglio}, {Davies}, {Mosser},
  {Girardi}, \& {Montalb{\'a}n}}]{Khan2018}
{Khan}, S., {Hall}, O.~J., {Miglio}, A., {et~al.} 2018, \apj, 859, 156

\bibitem[{{Kippenhahn} {et~al.}(1980){Kippenhahn}, {Ruschenplatt}, \&
  {Thomas}}]{Kippenhahn1980}
{Kippenhahn}, R., {Ruschenplatt}, G., \& {Thomas}, H.~C. 1980, \aap, 91, 175

\bibitem[{{Kippenhahn} \& {Weigert}(1994)}]{kippy}
{Kippenhahn}, R., \& {Weigert}, A. 1994, {Stellar Structure and Evolution}

\bibitem[{{Kirby} {et~al.}(2016){Kirby}, {Guhathakurta}, {Zhang}, {Hong},
  {Guo}, {Guo}, {Cohen}, \& {Cunha}}]{Kirby2016}
{Kirby}, E.~N., {Guhathakurta}, P., {Zhang}, A.~J., {et~al.} 2016, \apj, 819,
  135

\bibitem[{{Korre} {et~al.}(2019){Korre}, {Garaud}, \& {Brummell}}]{Korre2019}
{Korre}, L., {Garaud}, P., \& {Brummell}, N.~H. 2019, \mnras, 484, 1220

\bibitem[{{Kraft}(1994)}]{Kraft1994}
{Kraft}, R.~P. 1994, \pasp, 106, 553

\bibitem[{{Lagarde} {et~al.}(2011){Lagarde}, {Charbonnel}, {Decressin}, \&
  {Hagelberg}}]{Lagarde2011}
{Lagarde}, N., {Charbonnel}, C., {Decressin}, T., \& {Hagelberg}, J. 2011,
  \aap, 536, A28

\bibitem[{{Lagarde} {et~al.}(2012){Lagarde}, {Decressin}, {Charbonnel},
  {Eggenberger}, {Ekstr{\"o}m}, \& {Palacios}}]{Lagarde2012}
{Lagarde}, N., {Decressin}, T., {Charbonnel}, C., {et~al.} 2012, \aap, 543,
  A108

\bibitem[{{Lagarde} {et~al.}(2017){Lagarde}, {Robin}, {Reyl{\'e}}, \&
  {Nasello}}]{Lagarde2017}
{Lagarde}, N., {Robin}, A.~C., {Reyl{\'e}}, C., \& {Nasello}, G. 2017, \aap,
  601, A27

\bibitem[{{Lagarde} {et~al.}(2019){Lagarde}, {Reyl{\'e}}, {Robin},
  {Tautvai{\v{s}}ien{\.{e}}}, {Drazdauskas}, {Mikolaitis},
  {Minkevi{\v{c}}i{\={u}}t{\.{e}}}, {Stonkut{\.{e}}}, {Chorniy}, {Bagdonas},
  {Miglio}, {Nasello}, {Gilmore}, {Randich}, {Bensby}, {Bragaglia},
  {Flaccomio}, {Francois}, {Korn}, {Pancino}, {Smiljanic}, {Bayo}, {Carraro},
  {Costado}, {Jim{\'e}nez-Esteban}, {Jofr{\'e}}, {Martell}, {Masseron},
  {Monaco}, {Morbidelli}, {Sbordone}, {Sousa}, \& {Zaggia}}]{Lagarde2019}
{Lagarde}, N., {Reyl{\'e}}, C., {Robin}, A.~C., {et~al.} 2019, \aap, 621, A24

\bibitem[{{Lambert}(1981)}]{Lambert1981}
{Lambert}, D.~L. 1981, in Astrophysics and Space Science Library, Vol.~88,
  Physical Processes in Red Giants, ed. J.~{Iben}, I. \& A.~{Renzini}, 115--134

\bibitem[{{Langanke} \& {Mart{\'{\i}}nez-Pinedo}(2000)}]{Langanke2000}
{Langanke}, K., \& {Mart{\'{\i}}nez-Pinedo}, G. 2000, Nuclear Physics A, 673,
  481

\bibitem[{{Lattanzio} {et~al.}(2015){Lattanzio}, {Siess}, {Church}, {Angelou},
  {Stancliffe}, {Doherty}, {Stephen}, \& {Campbell}}]{Lattanzio2015}
{Lattanzio}, J.~C., {Siess}, L., {Church}, R.~P., {et~al.} 2015, \mnras, 446,
  2673

\bibitem[{{Ledoux}(1951)}]{Ledoux1951}
{Ledoux}, P. 1951, ApJ, 114, 373

\bibitem[{{Maeder} {et~al.}(2013){Maeder}, {Meynet}, {Lagarde}, \&
  {Charbonnel}}]{Maeder2013}
{Maeder}, A., {Meynet}, G., {Lagarde}, N., \& {Charbonnel}, C. 2013, \aap, 553,
  A1

\bibitem[{{Magrini} {et~al.}(2021{\natexlab{a}}){Magrini}, {Smiljanic},
  {Lagarde}, {Franciosini}, {Pasquini}, {Romano}, {Randich}, \&
  {Gilmore}}]{Magrini2021c}
{Magrini}, L., {Smiljanic}, R., {Lagarde}, N., {et~al.} 2021{\natexlab{a}}, The
  Messenger, 185, 18

\bibitem[{{Magrini} {et~al.}(2021{\natexlab{b}}){Magrini}, {Lagarde},
  {Charbonnel}, {Franciosini}, {Randich}, {Smiljanic}, {Casali}, {Viscasillas
  V{\'a}zquez}, {Spina}, {Biazzo}, {Pasquini}, {Bragaglia}, {Van der Swaelmen},
  {Tautvai{\v{s}}ien{\.{e}}}, {Inno}, {Sanna}, {Prisinzano}, {Degl'Innocenti},
  {Prada Moroni}, {Roccatagliata}, {Tognelli}, {Monaco}, {de Laverny},
  {Delgado-Mena}, {Baratella}, {D'Orazi}, {Vallenari}, {Gonneau}, {Worley},
  {Jim{\'e}nez-Esteban}, {Jofre}, {Bensby}, {Fran{\c{c}}ois}, {Guiglion},
  {Bayo}, {Jeffries}, {Binks}, {Gilmore}, {Damiani}, {Korn}, {Pancino},
  {Sacco}, {Hourihane}, {Morbidelli}, \& {Zaggia}}]{Magrini2021a}
{Magrini}, L., {Lagarde}, N., {Charbonnel}, C., {et~al.} 2021{\natexlab{b}},
  \aap, 651, A84

\bibitem[{{Majewski} {et~al.}(2017){Majewski}, {Schiavon}, {Frinchaboy},
  {Allende Prieto}, {Barkhouser}, {Bizyaev}, {Blank}, {Brunner}, {Burton},
  {Carrera}, {Chojnowski}, {Cunha}, {Epstein}, {Fitzgerald}, {Garc{\'{\i}}a
  P{\'e}rez}, {Hearty}, {Henderson}, {Holtzman}, {Johnson}, {Lam}, {Lawler},
  {Maseman}, {M{\'e}sz{\'a}ros}, {Nelson}, {Nguyen}, {Nidever}, {Pinsonneault},
  {Shetrone}, {Smee}, {Smith}, {Stolberg}, {Skrutskie}, {Walker}, {Wilson},
  {Zasowski}, {Anders}, {Basu}, {Beland}, {Blanton}, {Bovy}, {Brownstein},
  {Carlberg}, {Chaplin}, {Chiappini}, {Eisenstein}, {Elsworth}, {Feuillet},
  {Fleming}, {Galbraith-Frew}, {Garc{\'{\i}}a}, {Garc{\'{\i}}a-Hern{\'a}ndez},
  {Gillespie}, {Girardi}, {Gunn}, {Hasselquist}, {Hayden}, {Hekker}, {Ivans},
  {Kinemuchi}, {Klaene}, {Mahadevan}, {Mathur}, {Mosser}, {Muna}, {Munn},
  {Nichol}, {O'Connell}, {Parejko}, {Robin}, {Rocha-Pinto}, {Schultheis},
  {Serenelli}, {Shane}, {Silva Aguirre}, {Sobeck}, {Thompson}, {Troup},
  {Weinberg}, \& {Zamora}}]{Majewski2017}
{Majewski}, S.~R., {Schiavon}, R.~P., {Frinchaboy}, P.~M., {et~al.} 2017, \aj,
  154, 94

\bibitem[{{Martig} {et~al.}(2016){Martig}, {Fouesneau}, {Rix}, {Ness},
  {M{\'e}sz{\'a}ros}, {Garc{\'{\i}}a-Hern{\'a}ndez}, {Pinsonneault},
  {Serenelli}, {Silva Aguirre}, \& {Zamora}}]{Martig2016}
{Martig}, M., {Fouesneau}, M., {Rix}, H.-W., {et~al.} 2016, \mnras, 456, 3655

\bibitem[{{Masseron} \& {Gilmore}(2015)}]{MasseronGilmore2015}
{Masseron}, T., \& {Gilmore}, G. 2015, \mnras, 453, 1855

\bibitem[{{Meakin} \& {Arnett}(2007)}]{MeakinArnett2007}
{Meakin}, C.~A., \& {Arnett}, D. 2007, \apj, 667, 448

\bibitem[{{Nataf} {et~al.}(2013){Nataf}, {Gould}, {Pinsonneault}, \&
  {Udalski}}]{Nataf2013}
{Nataf}, D.~M., {Gould}, A.~P., {Pinsonneault}, M.~H., \& {Udalski}, A. 2013,
  \apj, 766, 77

\bibitem[{{Ness} {et~al.}(2016){Ness}, {Hogg}, {Rix}, {Martig}, {Pinsonneault},
  \& {Ho}}]{Ness2016}
{Ness}, M., {Hogg}, D.~W., {Rix}, H.-W., {et~al.} 2016, \apj, 823, 114

\bibitem[{{Nidever} {et~al.}(2015){Nidever}, {Holtzman}, {Allende Prieto},
  {Beland}, {Bender}, {Bizyaev}, {Burton}, {Desphande}, {Fleming},
  {Garc{\'{\i}}a P{\'e}rez}, {Hearty}, {Majewski}, {M{\'e}sz{\'a}ros}, {Muna},
  {Nguyen}, {Schiavon}, {Shetrone}, {Skrutskie}, {Sobeck}, \&
  {Wilson}}]{Nidever2015}
{Nidever}, D.~L., {Holtzman}, J.~A., {Allende Prieto}, C., {et~al.} 2015, \aj,
  150, 173

\bibitem[{{Oda} {et~al.}(1994){Oda}, {Hino}, {Muto}, {Takahara}, \&
  {Sato}}]{Oda1994}
{Oda}, T., {Hino}, M., {Muto}, K., {Takahara}, M., \& {Sato}, K. 1994, Atomic
  Data and Nuclear Data Tables, 56, 231

\bibitem[{{Palacios} {et~al.}(2006){Palacios}, {Charbonnel}, {Talon}, \&
  {Siess}}]{Palacios2006}
{Palacios}, A., {Charbonnel}, C., {Talon}, S., \& {Siess}, L. 2006, \aap, 453,
  261

\bibitem[{{Paxton} {et~al.}(2011){Paxton}, {Bildsten}, {Dotter}, {Herwig},
  {Lesaffre}, \& {Timmes}}]{paxton2011}
{Paxton}, B., {Bildsten}, L., {Dotter}, A., {et~al.} 2011, \apjs, 192, 3

\bibitem[{{Paxton} {et~al.}(2013){Paxton}, {Cantiello}, {Arras}, {Bildsten},
  {Brown}, {Dotter}, {Mankovich}, {Montgomery}, {Stello}, {Timmes}, \&
  {Townsend}}]{paxton2013}
{Paxton}, B., {Cantiello}, M., {Arras}, P., {et~al.} 2013, \apjs, 208, 4

\bibitem[{{Paxton} {et~al.}(2015){Paxton}, {Marchant}, {Schwab}, {Bauer},
  {Bildsten}, {Cantiello}, {Dessart}, {Farmer}, {Hu}, {Langer}, {Townsend},
  {Townsley}, \& {Timmes}}]{paxton2015}
{Paxton}, B., {Marchant}, P., {Schwab}, J., {et~al.} 2015, \apjs, 220, 15

\bibitem[{{Paxton} {et~al.}(2018){Paxton}, {Schwab}, {Bauer}, {Bildsten},
  {Blinnikov}, {Duffell}, {Farmer}, {Goldberg}, {Marchant}, {Sorokina},
  {Thoul}, {Townsend}, \& {Timmes}}]{Paxton2018}
{Paxton}, B., {Schwab}, J., {Bauer}, E.~B., {et~al.} 2018, \apjs, 234, 34

\bibitem[{{Paxton} {et~al.}(2019){Paxton}, {Smolec}, {Schwab}, {Gautschy},
  {Bildsten}, {Cantiello}, {Dotter}, {Farmer}, {Goldberg}, {Jermyn}, {Kanbur},
  {Marchant}, {Thoul}, {Townsend}, {Wolf}, {Zhang}, \& {Timmes}}]{paxton2019}
{Paxton}, B., {Smolec}, R., {Schwab}, J., {et~al.} 2019, \apjs, 243, 10

\bibitem[{{Pedersen} {et~al.}(2021){Pedersen}, {Aerts}, {P{\'a}pics},
  {Michielsen}, {Gebruers}, {Rogers}, {Molenberghs}, {Burssens}, {Garcia}, \&
  {Bowman}}]{Pedersen2021}
{Pedersen}, M.~G., {Aerts}, C., {P{\'a}pics}, P.~I., {et~al.} 2021, Nature
  Astronomy, arXiv:2105.04533

\bibitem[{{Pilachowski}(1986)}]{Pilachowski1986}
{Pilachowski}, C. 1986, \apj, 300, 289

\bibitem[{{Potekhin} \& {Chabrier}(2010)}]{Potekhin2010}
{Potekhin}, A.~Y., \& {Chabrier}, G. 2010, Contributions to Plasma Physics, 50,
  82

\bibitem[{{Rogers} \& {Nayfonov}(2002)}]{Rogers2002}
{Rogers}, F.~J., \& {Nayfonov}, A. 2002, \apj, 576, 1064

\bibitem[{{Saumon} {et~al.}(1995){Saumon}, {Chabrier}, \& {van
  Horn}}]{Saumon1995}
{Saumon}, D., {Chabrier}, G., \& {van Horn}, H.~M. 1995, \apjs, 99, 713

\bibitem[{{Sengupta} \& {Garaud}(2018)}]{SenguptaGaraud2018}
{Sengupta}, S., \& {Garaud}, P. 2018, \apj, 862, 136

\bibitem[{{Shetrone} {et~al.}(2019){Shetrone}, {Tayar}, {Johnson}, {Somers},
  {Pinsonneault}, {Holtzman}, {Hasselquist}, {Masseron}, {M{\'e}sz{\'a}ros},
  {J{\"o}nsson}, {Hawkins}, {Sobeck}, {Zamora}, \&
  {Garc{\'\i}a-Hern{\'a}ndez}}]{Shetrone2019}
{Shetrone}, M., {Tayar}, J., {Johnson}, J.~A., {et~al.} 2019, \apj, 872, 137

\bibitem[{{Stancliffe} {et~al.}(2009){Stancliffe}, {Church}, {Angelou}, \&
  {Lattanzio}}]{Stancliffe2009}
{Stancliffe}, R.~J., {Church}, R.~P., {Angelou}, G.~C., \& {Lattanzio}, J.~C.
  2009, \mnras, 396, 2313

\bibitem[{{Straus} {et~al.}(1976){Straus}, {Blake}, \& {Schramm}}]{Straus1976}
{Straus}, J.~M., {Blake}, J.~B., \& {Schramm}, D.~N. 1976, \apj, 204, 481

\bibitem[{{Sweigart} {et~al.}(1989){Sweigart}, {Greggio}, \&
  {Renzini}}]{Sweigart1989}
{Sweigart}, A.~V., {Greggio}, L., \& {Renzini}, A. 1989, \apjs, 69, 911

\bibitem[{{Timmes} \& {Swesty}(2000)}]{Timmes2000}
{Timmes}, F.~X., \& {Swesty}, F.~D. 2000, \apjs, 126, 501

\bibitem[{{Traxler} {et~al.}(2011){Traxler}, {Garaud}, \&
  {Stellmach}}]{Traxler2011}
{Traxler}, A., {Garaud}, P., \& {Stellmach}, S. 2011, \apjl, 728, L29

\bibitem[{{Ulrich}(1972)}]{Ulrich1972}
{Ulrich}, R.~K. 1972, \apj, 172, 165

\bibitem[{{Wilson} {et~al.}(2019){Wilson}, {Hearty}, {Skrutskie}, {Majewski},
  {Holtzman}, {Eisenstein}, {Gunn}, {Blank}, {Henderson}, {Smee}, {Nelson},
  {Nidever}, {Arns}, {Barkhouser}, {Barr}, {Beland}, {Bershady}, {Blanton},
  {Brunner}, {Burton}, {Carey}, {Carr}, {Colque}, {Crane}, {Damke}, {Davidson},
  {Dean}, {Di Mille}, {Don}, {Ebelke}, {Evans}, {Fitzgerald}, {Gillespie},
  {Hall}, {Harding}, {Harding}, {Hammond}, {Hancock}, {Harrison}, {Hope},
  {Horne}, {Karakla}, {Lam}, {Leger}, {MacDonald}, {Maseman}, {Matsunari},
  {Melton}, {Mitcheltree}, {O'Brien}, {O'Connell}, {Patten}, {Richardson},
  {Rieke}, {Rieke}, {Roman-Lopes}, {Schiavon}, {Sobeck}, {Stolberg}, {Stoll},
  {Tembe}, {Trujillo}, {Uomoto}, {Vernieri}, {Walker}, {Weinberg}, {Young},
  {Anthony-Brumfield}, {Bizyaev}, {Breslauer}, {De Lee}, {Downey}, {Halverson},
  {Huehnerhoff}, {Klaene}, {Leon}, {Long}, {Mahadevan}, {Malanushenko},
  {Nguyen}, {Owen}, {S{\'a}nchez-Gallego}, {Sayres}, {Shane}, {Shectman},
  {Shetrone}, {Skinner}, {Stauffer}, \& {Zhao}}]{Wilson2019}
{Wilson}, J.~C., {Hearty}, F.~R., {Skrutskie}, M.~F., {et~al.} 2019, \pasp,
  131, 055001

\bibitem[{{Zahn}(1991)}]{Zahn1991}
{Zahn}, J.~P. 1991, \aap, 252, 179

\bibitem[{{Zamora} {et~al.}(2015){Zamora}, {Garc{\'{\i}}a-Hern{\'a}ndez},
  {Allende Prieto}, {Carrera}, {Koesterke}, {Edvardsson}, {Castelli}, {Plez},
  {Bizyaev}, {Cunha}, {Garc{\'{\i}}a P{\'e}rez}, {Gustafsson}, {Holtzman},
  {Lawler}, {Majewski}, {Manchado}, {M{\'e}sz{\'a}ros}, {Shane}, {Shetrone},
  {Smith}, \& {Zasowski}}]{Zamora2015}
{Zamora}, O., {Garc{\'{\i}}a-Hern{\'a}ndez}, D.~A., {Allende Prieto}, C.,
  {et~al.} 2015, \aj, 149, 181

\bibitem[{{Zasowski} {et~al.}(2017){Zasowski}, {Cohen}, {Chojnowski},
  {Santana}, {Oelkers}, {Andrews}, {Beaton}, {Bender}, {Bird}, {Bovy},
  {Carlberg}, {Covey}, {Cunha}, {Dell'Agli}, {Fleming}, {Frinchaboy},
  {Garc{\'{\i}}a-Hern{\'a}ndez}, {Harding}, {Holtzman}, {Johnson}, {Kollmeier},
  {Majewski}, {M{\'e}sz{\'a}ros}, {Munn}, {Mu{\~n}oz}, {Ness}, {Nidever},
  {Poleski}, {Rom{\'a}n-Z{\'u}{\~n}iga}, {Shetrone}, {Simon}, {Smith},
  {Sobeck}, {Stringfellow}, {Szigeti{\'a}ros}, {Tayar}, \&
  {Troup}}]{Zasowski2017}
{Zasowski}, G., {Cohen}, R.~E., {Chojnowski}, S.~D., {et~al.} 2017, \aj, 154,
  198

\end{thebibliography}

\label{lastpage}
\end{document}